\documentclass[a4paper,12pt]{article}

\usepackage{graphicx}
\usepackage{amssymb,amsmath,subfloat}
\usepackage{amsthm}
\usepackage{setspace}
\usepackage{geometry}
\usepackage{multirow}
\usepackage{caption}
\DeclareCaptionLabelFormat{cont}{#1~#2\alph{ContinuedFloat}}
\begin{document}


\begin{center}
\textbf{\Large Vibrational dynamics of phenylphenanthrenes with phenyl group at different positions}
~\\
~\\
Rashmi Singh and Shantanu Rastogi*
\\
Department of Physics, D.D.U. Gorakhpur University, Gorakhpur, India - 273009
\\
* e-mail:shantanu\_r@hotmail.com

\end{center}

\begin{abstract}

Polycyclic aromatic hydrocarbons (PAHs) appear to be ubiquitous in terrestrial atmosphere as well as in the interstellar medium (ISM). In the astrophysical context presence of PAHs is interpreted by the observations of mid-IR emission bands at 3.3, 6.2, 7.7, 8.6, 11.2 and 12.6 micron (3030, 1610, 1300, 1160, 890 and 790~cm$^{-1}$). The profile of these emission bands exhibit definite variations from source to source, which are explained due to sources having different types of PAHs in different ionization states. The model spectra of mixtures of PAHs show good fit for most bands except for the 6.2 micron feature. Thus a wider variety of PAHs need to be considered for this feature. PAH formation in circumstellar medium of planetary nebulae and dense clouds indicate formation of phenyl radical and phenyl substituted PAHs. To understand the modifications in infrared lineaments upon phenyl substitution and also due to substitution at different positions, study is made on five possible isomers of phenylphenanthrene. The features that may account for possible phenylphenanthrenes in protoplanetary nebulae atmospheres are also discussed. Possible contribution of phenylphenanthrenes towards the 6.2 micron feature and other astrophysical emission bands is assessed.

\end{abstract}

{\bf keywords:}
PAH; Aromatic Infrared Bands; Interstellar molecules; DFT calculations



\section{Introduction}

Among the carbonaceous molecules polycyclic aromatic hydrocarbons (PAHs) are ubiquitously found in terrestrial and in extraterrestrial space. In the Earth's atmosphere PAHs get introduced via any incomplete combustion process \cite{PAH-health}. In space the circumstellar shells of carbon rich stars in late stages of evolution have similar but low-density combustion like atmospheres \cite{Cherchneff2011}. PAH formation in planetary nebulae environments are envisaged through a bottom-up reaction approach \cite{Tielens2008}. Being very stable molecules, the PAHs seem to survive in harsh astrophysical conditions and are also assumed to form through defragmentation or surface etching of grains in a top-down approach \cite{Merino2014}. On Earth PAHs are well identified, several of these being carcinogenic and health hazard \cite{PAH-pollu}. Whereas, in space PAHs are interpreted on the basis of their infrared emission bands with no specific PAH identification possible.

Astrophysical infrared PAH emission bands, referred to as aromatic infrared bands (AIBs), appear at 3.3, 6.2, 7.7, 8.6, 11.2 and 12.6~$\mu$m (3030, 1610, 1300, 1160, 890 and 790~cm$^{-1}$) \cite{Cohen89, ISO96, Lutz98, Peeters2002}. These features result from mixing of emissions from a population of PAHs possible in that environment. The variations in these features are correlated with the type of astrophysical object and thus indicate different types of PAHs in different environments \cite{Pathak2008}. Study of infrared spectroscopic properties of different types of PAHs in different ionization states is, therefore, important to decipher astrophysical observations. This in turn helps in improving understanding of the physical and chemical evolution of various objects.

The astrophysical PAH formation scenario indicates phenyl radical as an important intermediate and may play a major role in chemistry of the interstellar medium (ISM) \cite{McMahon2003}. Modelling protoplanetary nebula conditions, of dense-warm gas irradiated by a strong UV field, has shown to produce benzene \cite{Woods2002}. Laboratory study of shock against benzene yield phenanthrene and phenylphenanthrenes \cite{Mimura2003}. Similar conditions are there in the expanding envelopes of protoplanetary nebulae, suggesting phenylphenanthrenes in the ISM. The presence of phenyl as a side group in aliphatic compounds and in complex systems show a medium intense mode close to the 6.2~$\mu$m AIB \cite{Aspartate96, Ngaojampa2015}. This AIB remains unexplained by considering plain PAHs \cite{Langhoff96, Pathak2005, Pathak2006, Pathak2007}. In this communication vibrational dynamics study of all five possible isomers of phenylphenanthrenes, shown in Figure~\ref{Fig1-4}, is performed. The paper is organized as follows, Section~1 gives the background information of the problem and covers the motivation behind the current work, Section~2 details the theoretical methods and results of computation, Section~3 presents vibrational analysis to discuss the modifications in the infrared features due to different attachment position of phenyl group and due to ionization, Section~4 presents the astrophysical aspects and the possibility of incorporation of phenylphenanthrenes in explaining astrophysical bands and Section~5 provides the concluding highlights of the work and gives pointers for possible future study.

\section{Theoretical Computation}

The type of PAHs that are possible in diverse conditions of the ISM may not all be available terrestrially for laboratory study. Quantum chemical calculations provide a useful tool to simulate infrared spectrum of such PAHs and consider their contribution towards AIBs \cite{Langhoff96, Pathak2005, Pathak2006, Pathak2007, Bauschlicher2008, Pathak2008, Rosenberg2014}. Even where the sample PAH is available, correct spectra of isolated molecule, as is possible in the ISM, is difficult in the laboratory. Theoretical computations are capable of providing suitable information for such systems.

As phenanthrene and phenylphenanthrenes are possible in shock regions \cite{Mimura2003} it is important to study their vibrational modes. In laboratory the four isomers 1--, 2--, 3-- and 4--phenylphenanthrene are obtained by photocyclization of phenylstilbenes \cite{Dickerman74}, while 9--phenylphenanthrene is obtainable via photocatalytic benzannulation \cite{Chatterjee2017}. The vibrational spectroscopic features of phenylphenanthrenes are not available in literature so a theoretical study of their infrared spectra is performed. The GAMESS ab initio program \cite{Gamess1993} is used to obtain optimized molecular structure and infrared vibrational frequencies. The calculations are performed using density functional theory (DFT) with B3LYP functionals in combination with the 6-31G(d) basis set. Study is performed on both neutrals and cations as molecules in the ISM are likely to be also in ionized form.

The phenyl moiety is out-of-plane with respect to the phenanthrene unit due to the steric hinderance from nearby hydrogens. Both the structure and stability of phenylphenanthrene isomers depend on the hydrogen placement with respect to the attached phenyl ring. The stability can be interpreted on the basis of the sharpness of minimized energy with respect to the out-of-plane dihedral angles of the phenyl moiety. The plot of energy vs. angle of phenyl torsion are shown in Figure~\ref{Fig2-4A}a for 1--, 4-- and 9--phenylphenanthrene, where the minima is at $\sim$55$^\circ$ and there is little increase in energy beyond the minima. For 2-- and 3--phenylphenanthrene, shown in Figure~\ref{Fig2-4B}b, the torsion angle minima are at $\sim$40$^\circ$. The environment of the phenyl group is similar in 2-- and 3--phenylphenanthrene with symmetric placement of hydrogen atoms. The placement of hydrogens in 1--, 4-- and 9--phenylphenanthrene is not symmetric (Figure~\ref{Fig1-4}).

The final optimized phenyl torsion angles are given in Table~\ref{tab1-4} along with the computed energies for both neutrals and cations. Upon ionization the torsion angle is reduced in all the isomers. The structural similarity of 2-- and 3--phenylphenanthrene, with respect to hydrogens surrounding the phenyl group, is reflected in their nearly equal minimized energies. These two conformers also have lowest energy indicating that symmetric hydrogen placement brings more stability to these isomers. The study of distribution and occurrence of phenylated aromatics in geological samples shows the presence of phenylphenanthrenes \cite{Rospondek2009, Li2012} wherein 2-- and 3--phenylphenanthrene are more prevalent than other isomers \cite{Li2012}. The highest optimized energy is that of 4--phenylphenanthrene, which is the most sterically hindered isomer having phenyl group in the bay region of phenanthrene.

The vibrational modes are computed using the optimized geometry. The frequencies obtained are usually overestimated and need to be scaled. The scaling procedure and factors depend upon the level of theory and basis sets \cite{Maurya2012}. The 6-31G(d) basis set is used in the current computations for phenylphenanthrenes and the two level frequency scaling \cite{Maurya2012} is applied. The scale factor for C-H stretching vibrations computed around 3000~cm$^{-1}$ is 0.9603 and that for the 1800~-~500~cm$^{-1}$ frequency range it is 0.9697.

\section{Vibrational analysis}

The scaled frequencies and the corresponding infrared intensities are used to simulate absorption spectrum of each isomer. The peaks are considered having Lorentzian profile with 5~cm$^{-1}$ FWHM and the intensities are taken relative to the most intense mode. These spectra are shown in Figures~\ref{Fig3-4}~--~\ref{Fig7-4} for the frequency range 1800~-~500~cm$^{-1}$, wherein vertical dotted lines show the position of 6.2 and 7.7~$\mu$m AIB. The C-H stretch mode frequencies, around 3000~cm$^{-1}$, fall close together and combine into a complex feature as shown in Figure~\ref{Fig8-4}. The frequencies and corresponding infrared intensities for each molecule and its cation are given in Tables~\ref{tab2-4}~--~\ref{tab6-4}, where only the modes having relative intensity $\ge$~0.01 are given. Just like in plain PAHs \cite{Langhoff96, Pathak2005, Pathak2006} neutral phenylphenanthrenes also have intense C-H stretch mode peaks that fall to very low intensities upon ionization. Ionization also leads to strong peaks in the 1600~--~1100~cm$^{-1}$ region corresponding to C-C stretch and in-plane bend modes.

In neutral phenylphenanthrenes prominent bands are related to motion of hydrogens, i.e. C-H stretching vibration in the 3000~cm$^{-1}$ region and C-H out-of-plane wag motion in 600~--~900~cm$^{-1}$ range. Although the hydrogens are attached to similar aromatic carbons yet even slight difference in non-bonded environment may change the vibrational frequency and intensity of the modes. In PAHs the hydrogen motions can be different according to it being attached to a ring that has quarto (four), trio (three), duo (two) or solo (one) peripheral hydrogens \cite{Bellamy80, Hudgins99}. In addition to a duo and two quarto hydrogens, there is also a bay region in phenanthrene \cite{Maltseva2016}, which when substituted with a phenyl group forms 4--phenylphenanthrene. Important modes in the five isomers are shown in Table~\ref{tab7-4}.

The features of phenanthrene spectrum \cite{Pathak2005} are only slightly disturbed by the addition of phenyl group. The phenyl group substitution induces its own bands and as the symmetry of phenanthrene is broken a few more modes get active. The phenyl group C-H wag is observed at $\sim$695~cm$^{-1}$ in all isomers with nearly equivalent intensity. The most intense single quarto C-H wag peak of phenanthrene \cite{Pathak2005} at $\sim$740~cm$^{-1}$ is two peaks in phenylphenanthrenes. In 1--phenylphenanthrene these are very close peaks of nearly equal strength, arising due to the breaking of phenanthrene symmetry on phenyl addition. In the other isomers the two peaks are about 20~cm$^{-1}$ apart with the lower frequency quarto C-H wag mode being stronger. The higher frequency mode is due to the C-H wag in the phenyl group. Another strong C-H wag mode of phenanthrene at 825~cm$^{-1}$ \cite{Pathak2005} is similar in 3-- and 4--phenylphenanthrene but it appears at 803~cm$^{-1}$ in 1-- and 2--phenylphenanthrene. In 9--phenylphenanthrene there is altogether a different mode at 769~cm$^{-1}$.

The strongest features in the neutral molecules are due to C-H stretch modes, the last column in Table~\ref{tab7-4}. For all isomers the most intense mode is at around 3070~cm$^{-1}$, due mainly to the C-H stretching in the phenanthrene unit. The strong higher frequency mode at $\sim$3090~cm$^{-1}$ is due to the C-H stretch of the two bay hydrogens. In 4--phenylphenanthrene this mode is absent, possibly as one bay hydrogen is substituted by the phenyl group.

Among other significant modes in the neutrals are planar ring deformation modes listed in Table~\ref{tab8-4}. Two peaks at around 1490 and 1455~cm$^{-1}$ related to C-C stretch and C-H in-plane bend modes are common in all isomers. The mode close to 1500~cm$^{-1}$ is reminiscent of the peak in phenanthrene \cite{Pathak2005}. 

In cations the intensity of C-H stretch modes is extremely small. The scale is magnified in the right column of Figure~\ref{Fig8-4} to see the features in cations. In addition the intensity of several modes in the 1600~--~1100~cm$^{-1}$ region is enhanced. The most intense feature is around 1590~cm$^{-1}$, due to C-C stretch in the phenyl group, in all isomers except 3-- and 4--phenylphenanthrene. In 3--phenylphenanthrene the most intense mode is at 1556~cm$^{-1}$ and in 4--phenylphenanthrene it is at 1400~cm$^{-1}$. The C-H out-of-plane wag modes get slightly shifted in the cations, cations, while their absolute intensity is not much changed. Their relative intensity appears to be small as the reference intensity of C-C stretch mode is very high..

\section{Astrophysical Aspects}

Study of shock against benzene in laboratory has shown to yield phenanthrene and phenylphenanthrenes \cite{Mimura2003}. Similar conditions are there in the expanding envelopes of protoplanetary nebulae, suggesting possibility of reactions leading to growth of phenylphenanthrenes in the ISM. The identification of phenylphenanthrenes in geological samples from different eras \cite{Rospondek2009, Li2012} also indicate the ease of growth for phenylated PAHs in carbonaceous environments. Although the astrophysical AIBs get explained through emission models using plain unsubstituted PAHs \cite{Pathak2008, Rosenberg2014}, yet some features like the 6.2~$\mu$m band remain unexplained. Phenyl substitution on phenanthrene does make some modes active in this frequency range.

In neutral phenylphenanthrenes the C-C stretch modes (Table~\ref{tab8-4}) fall right in the position of Class `A' AIB at 6.2~$\mu$m (1610~cm$^{-1}$) \cite{Peeters2002}, but with moderate intensity. The cations of 1--, 2-- and 9--phenylphenanthrene do have very strong feature at 6.3~$\mu$m (1590~cm$^{-1}$) that may explain the Class `C' AIB feature observed along protoplanetary nebulae type evolved stars \cite{Peeters2002}.

Other prominent features that appear in neutral phenylphenanthrenes include modes around 1495 and 1455~cm$^{-1}$ (6.7 and 6.9~$\mu$m). These fall close to AIB sub features observed along some objects. A feature at 6.66~$\mu$m is observed along IRAS 18434-0242 \cite{Peeters99} and in H2 and seyfert~2 galaxies \cite{Laureijs2000}. The 6.9~$\mu$m feature is seen in proto planetary nebulae \cite{Hrivnak2000} and in cool objects \cite{Buss90}. Incorporating suitable phenylphenanthrenes in emission modelling for specific objects is strongly indicated. Modelling of the 7.7~$\mu$m feature in protoplanetary nebulae point towards large PAHs in these objects \cite{Pathak2008}.

\section{Conclusion and Future Scope}

All the isomers of phenylphenanthrene in neutral form have a feature close to 6.2~$\mu$m but with moderate intensity. It is desirable to incorporate them for modelling the emission of specific objects. The possible objects appear to be those with cool atmospheres as around proto planetary nebulae and stars in the late stage of evolution. The Phenyl substituted large PAHs and multi-phenyl PAHs also need to be studied for possible improvement in the models and to give better insight into the physical and chemical conditions around such stars.


\clearpage
\begin{figure}
\centerline{\includegraphics[width=1.0\textwidth]{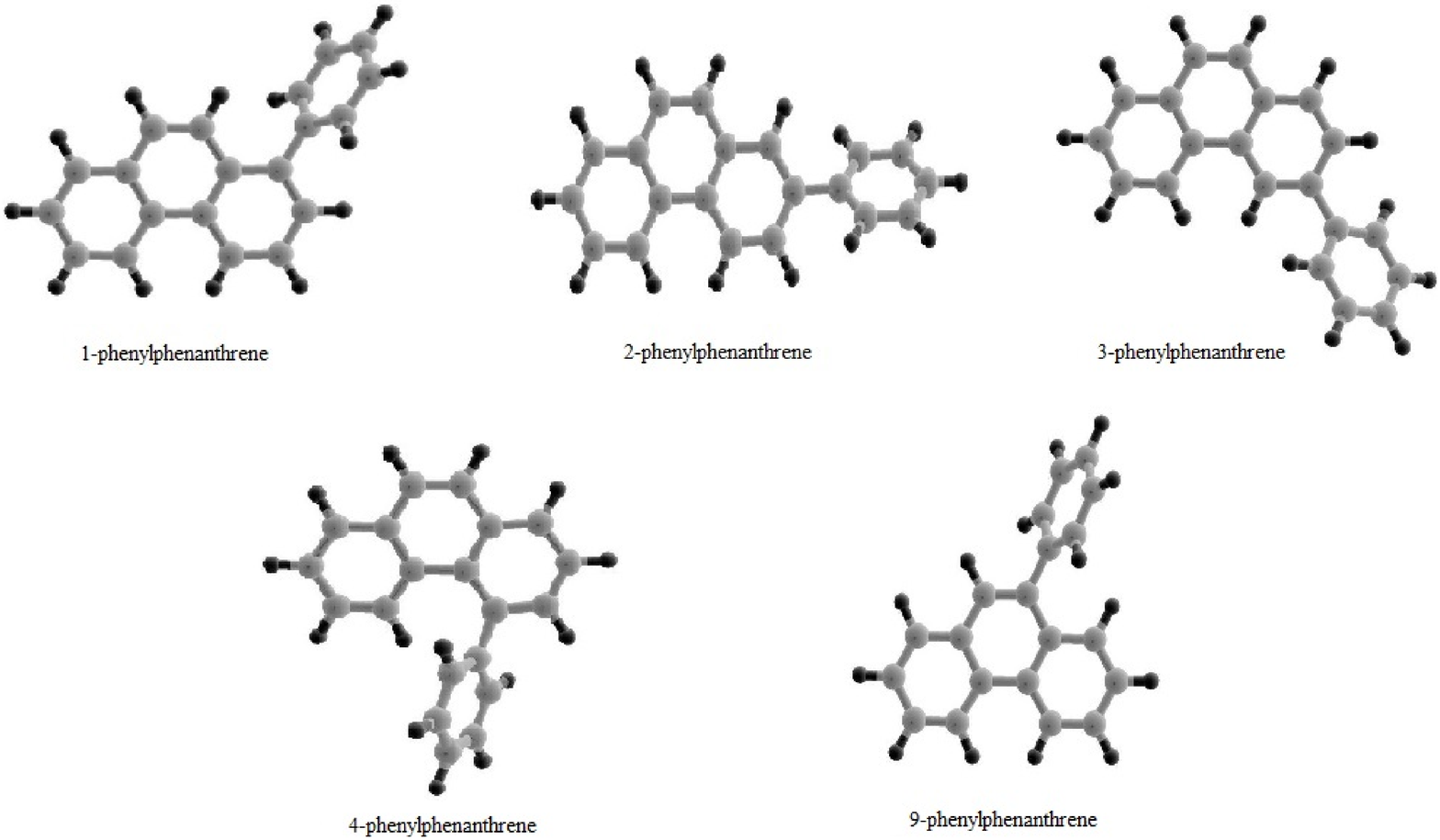}}
\caption{Molecular structures of five phenylphenanthrene isomers.}
\label{Fig1-4}
\end{figure}
\clearpage

\clearpage
\begin{figure}
     \ContinuedFloat*
    \centerline{\includegraphics[width=0.5\textwidth]{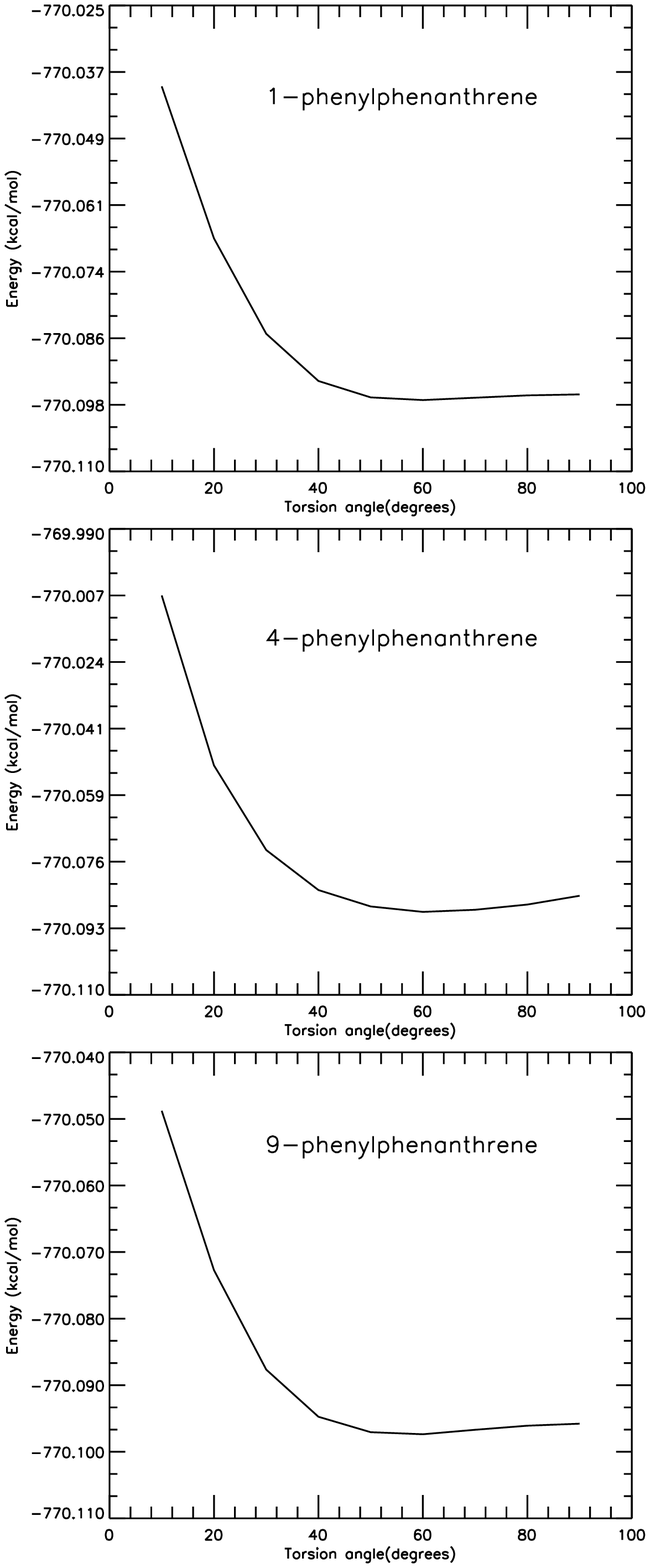}}
    \caption{Variation of optimization energy with torsion angle of the phenyl group substituted in asymmetric H position.}
    \label{Fig2-4A}
\end{figure}

\clearpage
\begin{figure}
     \ContinuedFloat
    \centerline{\includegraphics[width=0.5\textwidth]{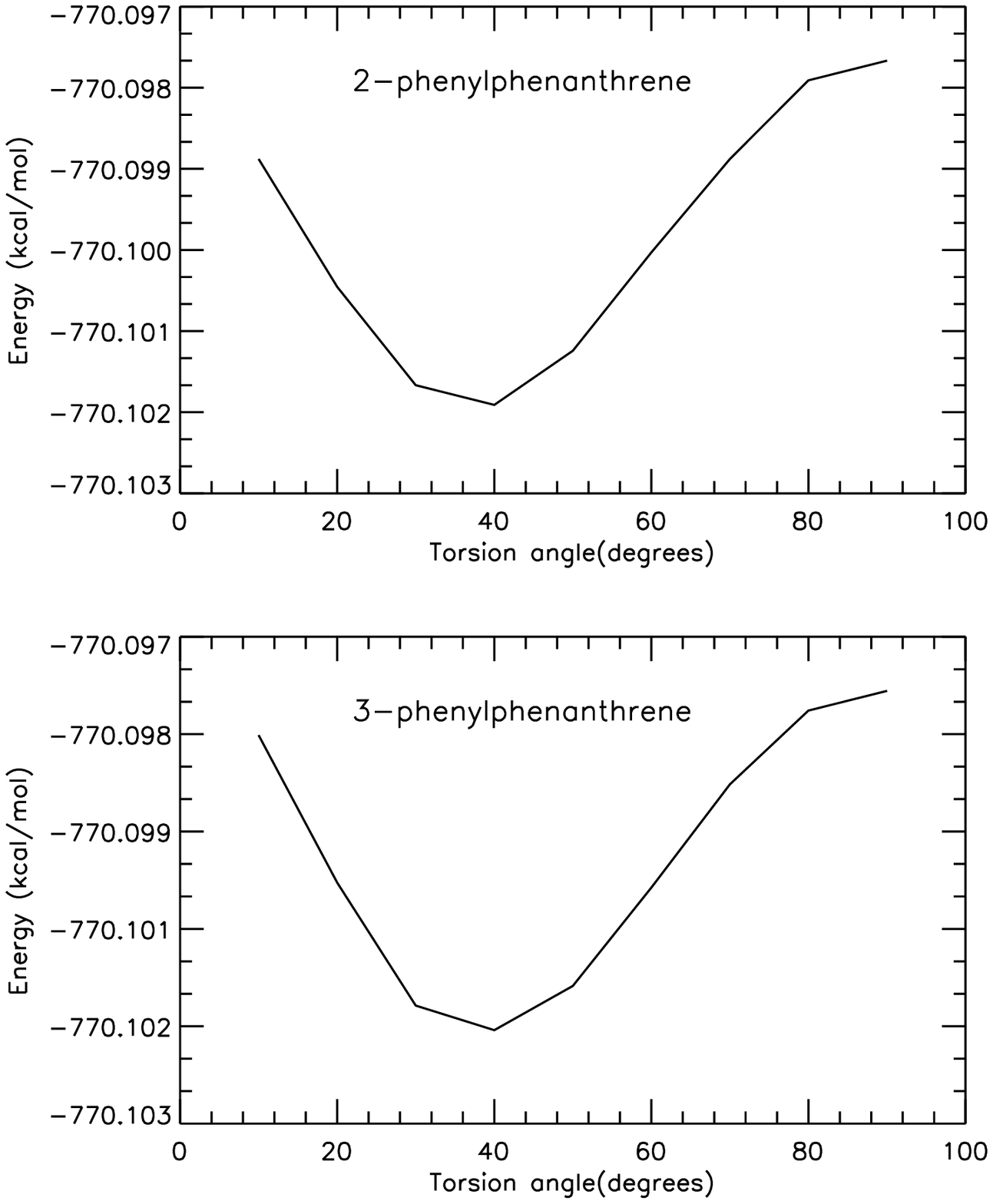}}
    \caption{Variation of optimization energy with torsion angle of the phenyl group substituted in symmetric H position.}
    \label{Fig2-4B}
\end{figure}

\clearpage
\begin{figure}
\centerline{\includegraphics[width=0.8\textwidth]{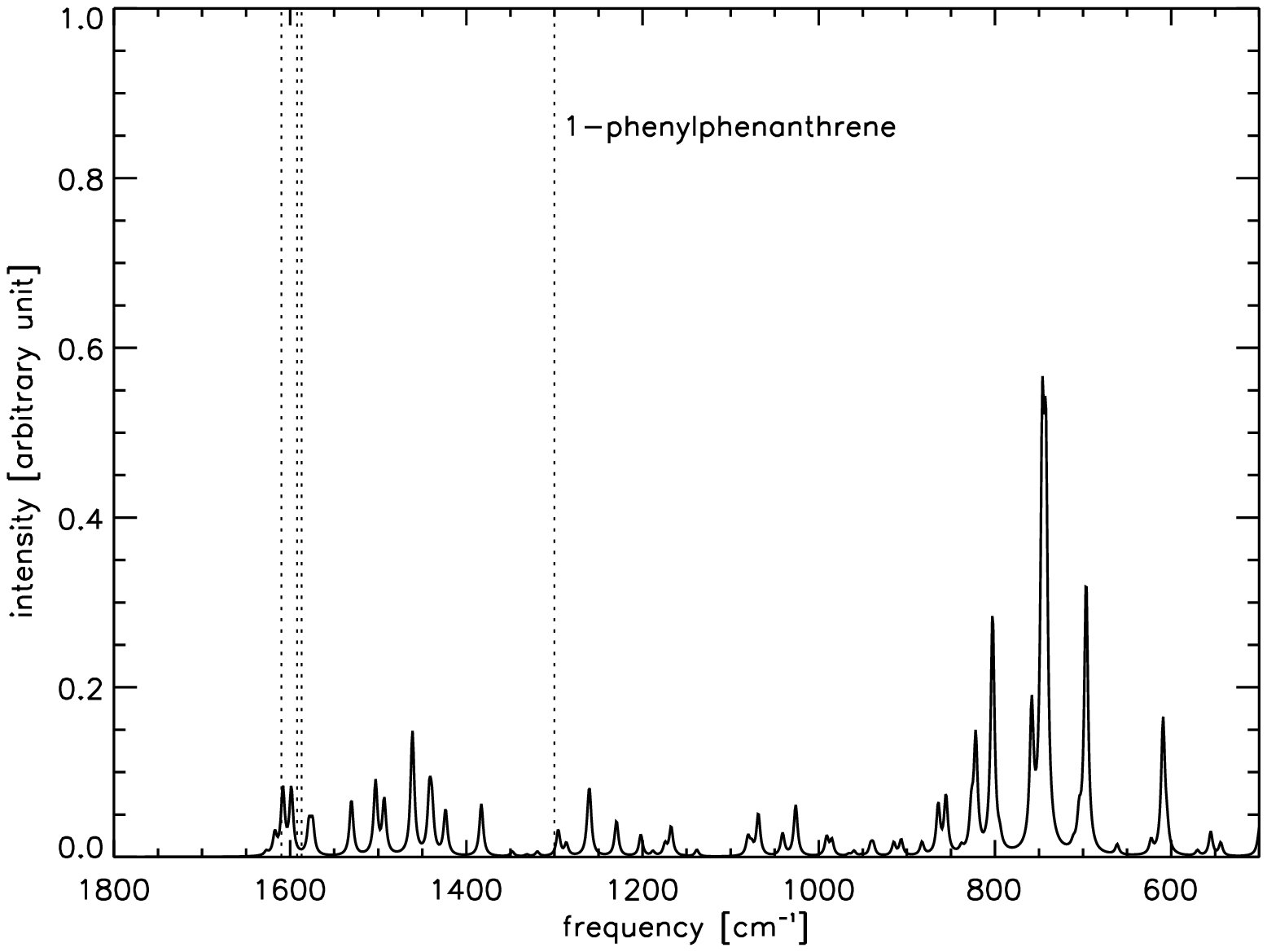}}
\centerline{\includegraphics[width=0.8\textwidth]{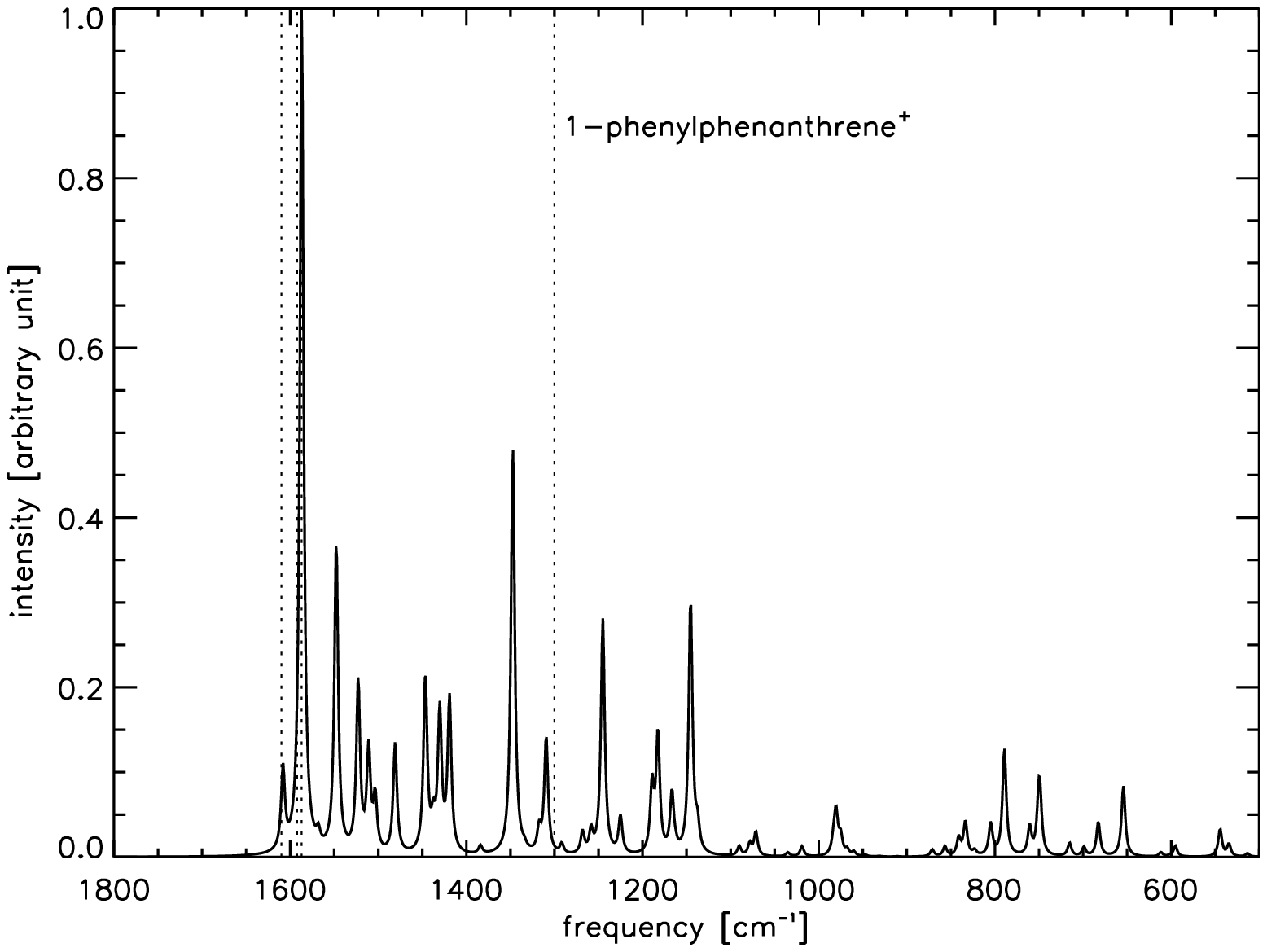}}
\caption{Infrared spectra of 1--phenylphenanthrene neutral and cation}
\label{Fig3-4}
\end{figure}

\clearpage
\begin{figure}
\centerline{\includegraphics[width=0.8\textwidth]{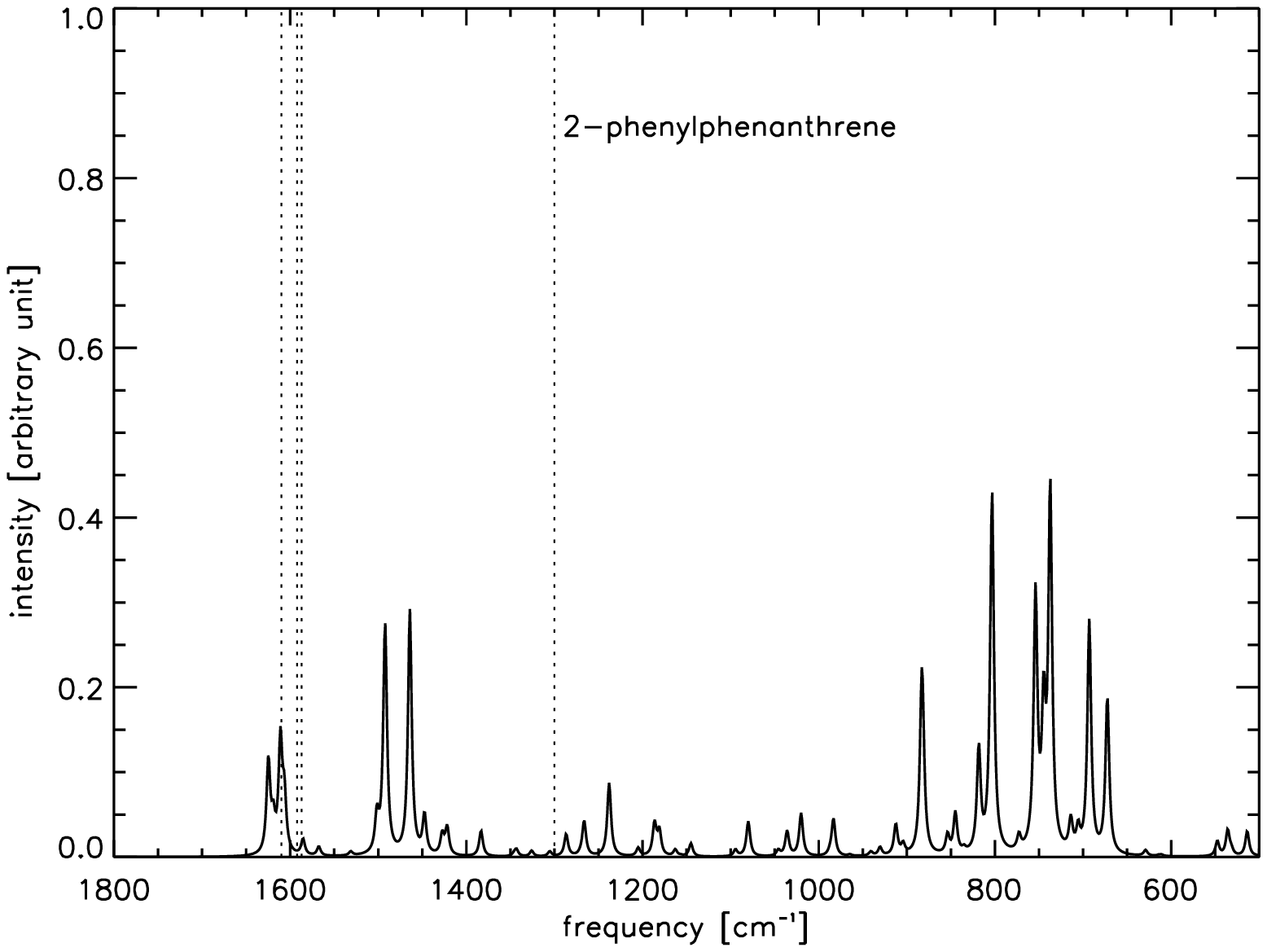}}
\centerline{\includegraphics[width=0.8\textwidth]{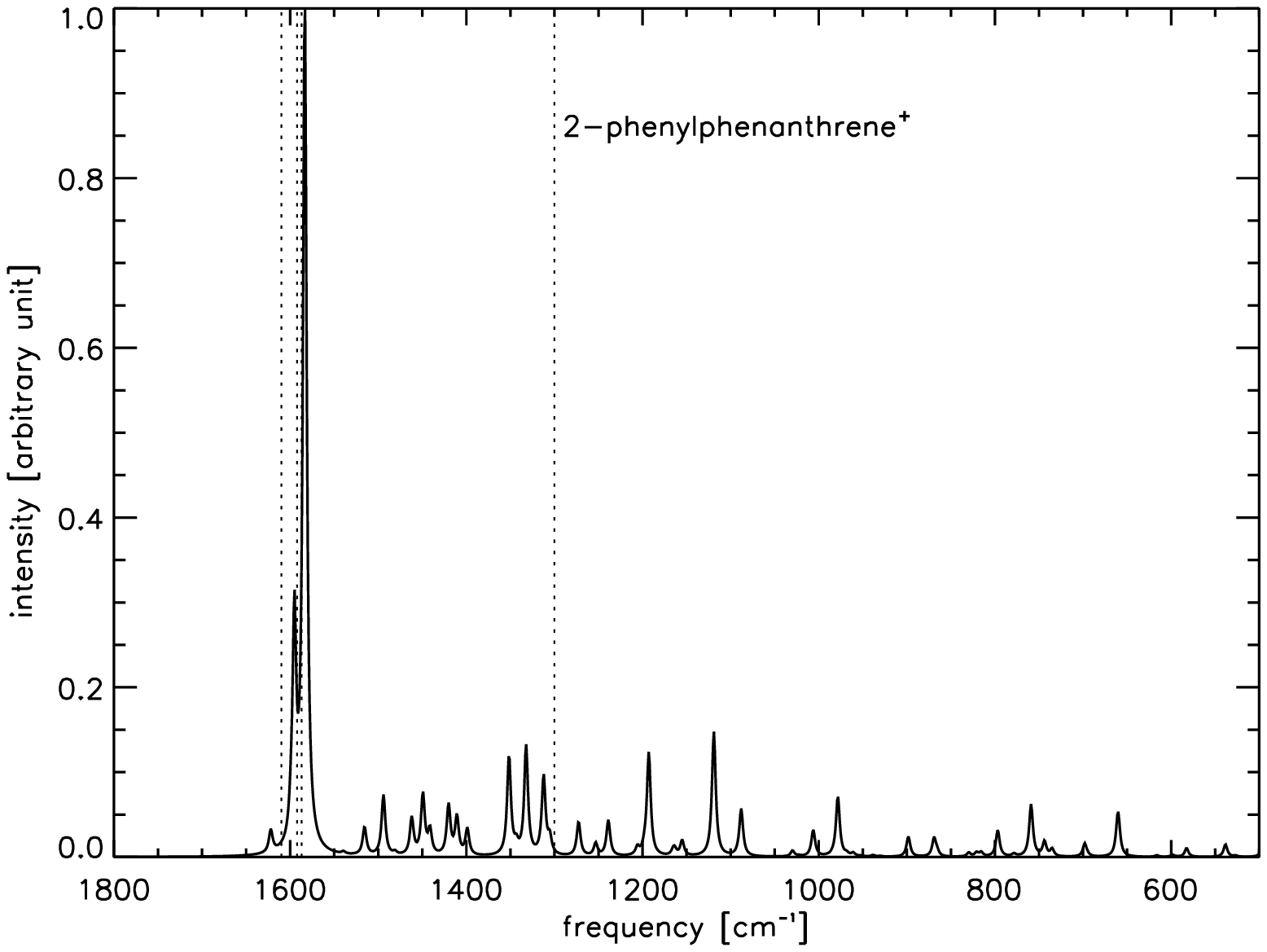}}
\caption{Infrared spectra of 2--phenylphenanthrene neutral and cation}
\label{Fig4-4}
\end{figure}

\clearpage
\begin{figure}
\centerline{\includegraphics[width=0.8\textwidth]{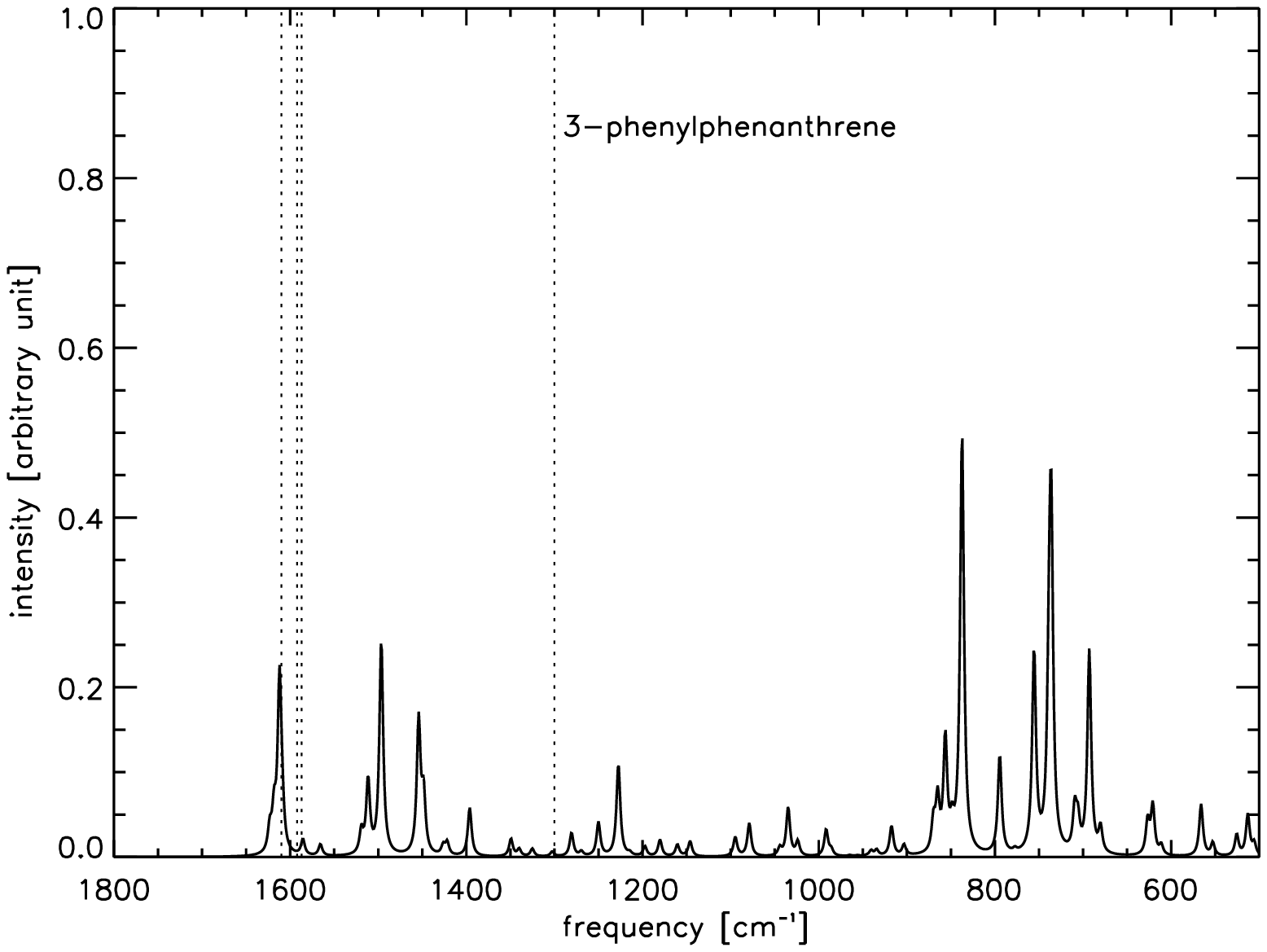}}
\centerline{\includegraphics[width=0.8\textwidth]{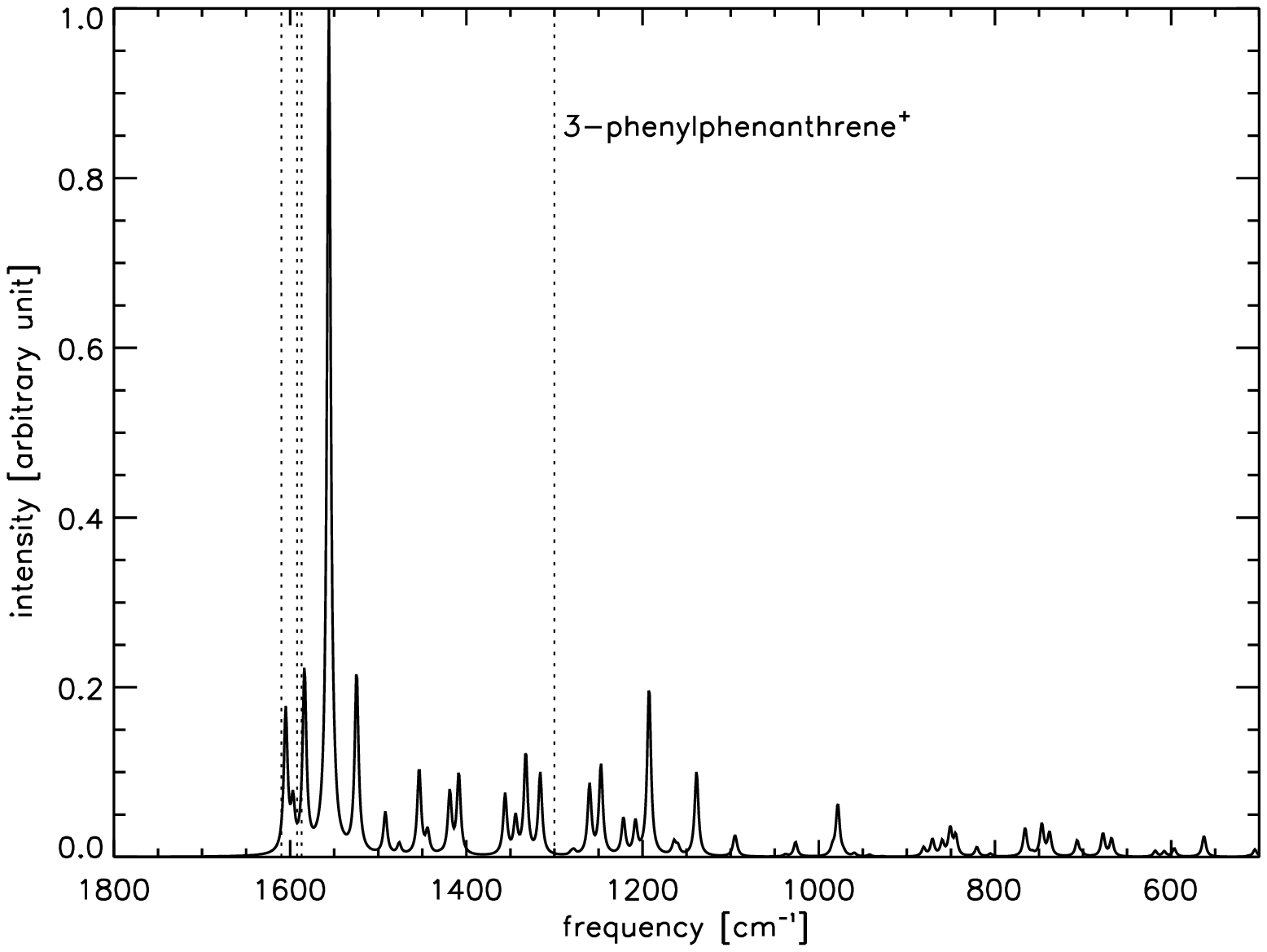}}
\caption{Infrared spectra of 3--phenylphenanthrene neutral and cation}
\label{Fig5-4}
\end{figure}

\clearpage
\begin{figure}
\centerline{\includegraphics[width=0.8\textwidth]{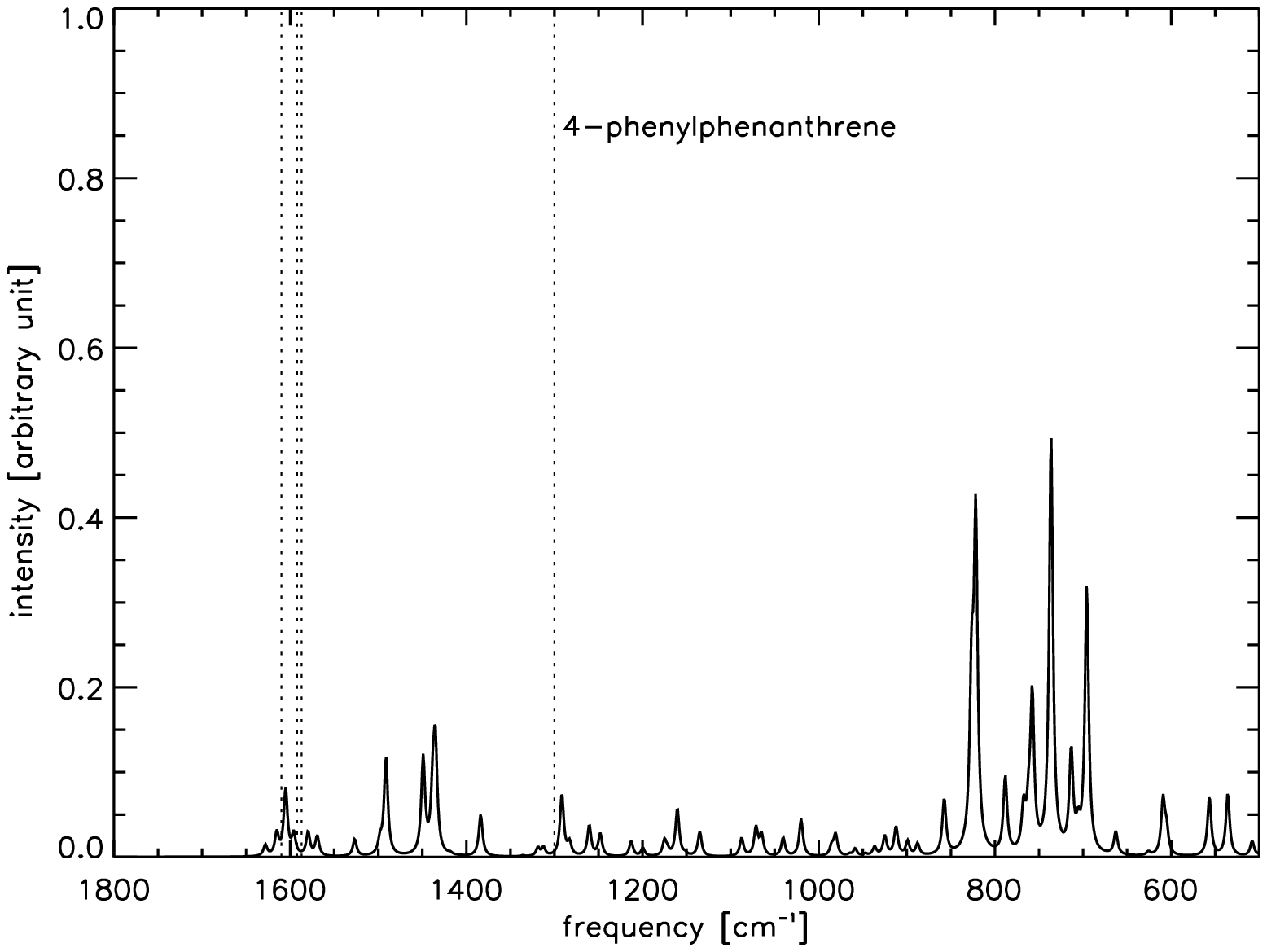}}
\centerline{\includegraphics[width=0.8\textwidth]{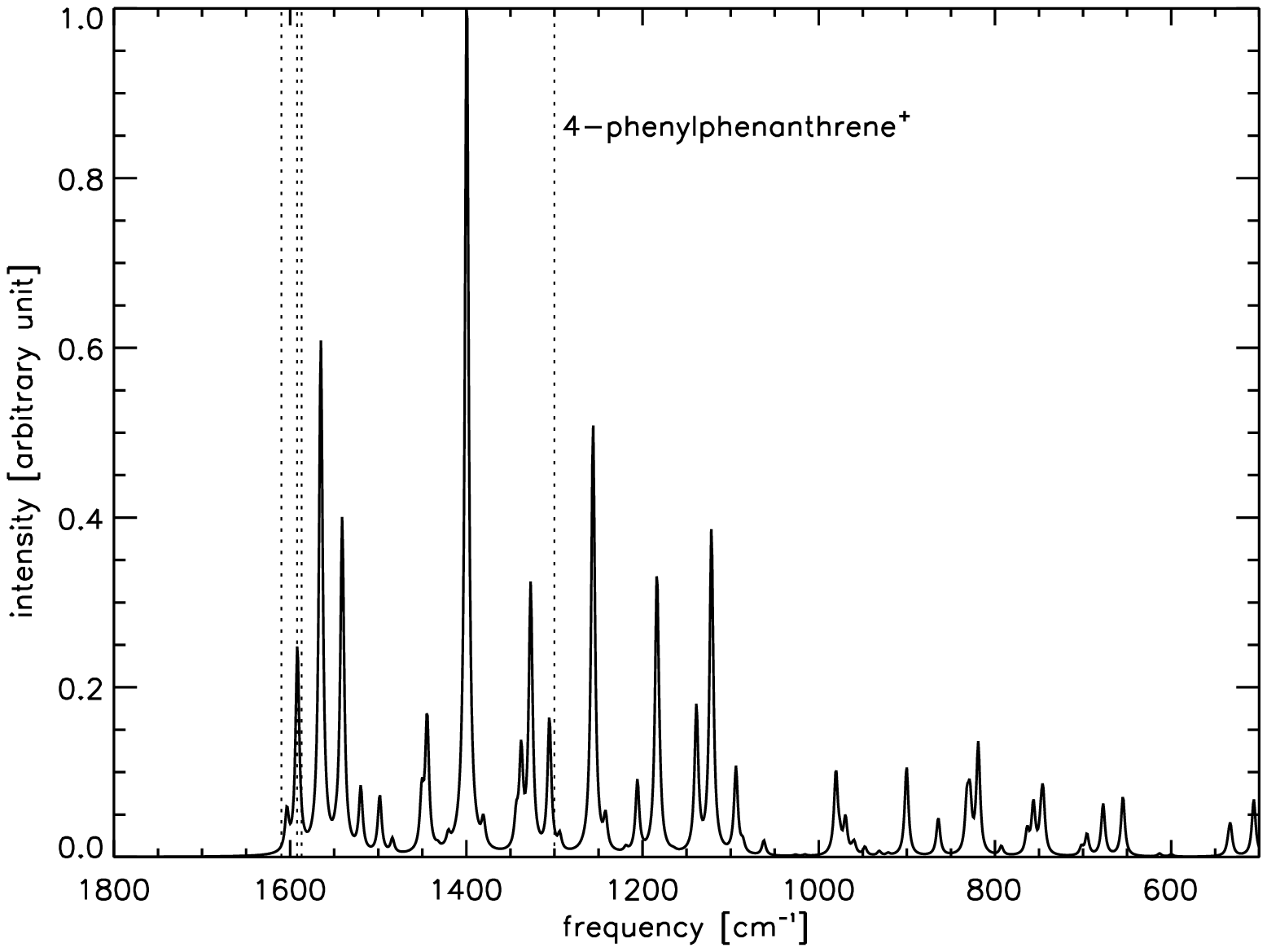}}
\caption{Infrared spectra of 4--phenylphenanthrene neutral and cation}
\label{Fig6-4}
\end{figure}

\clearpage
\begin{figure}
\centerline{\includegraphics[width=0.8\textwidth]{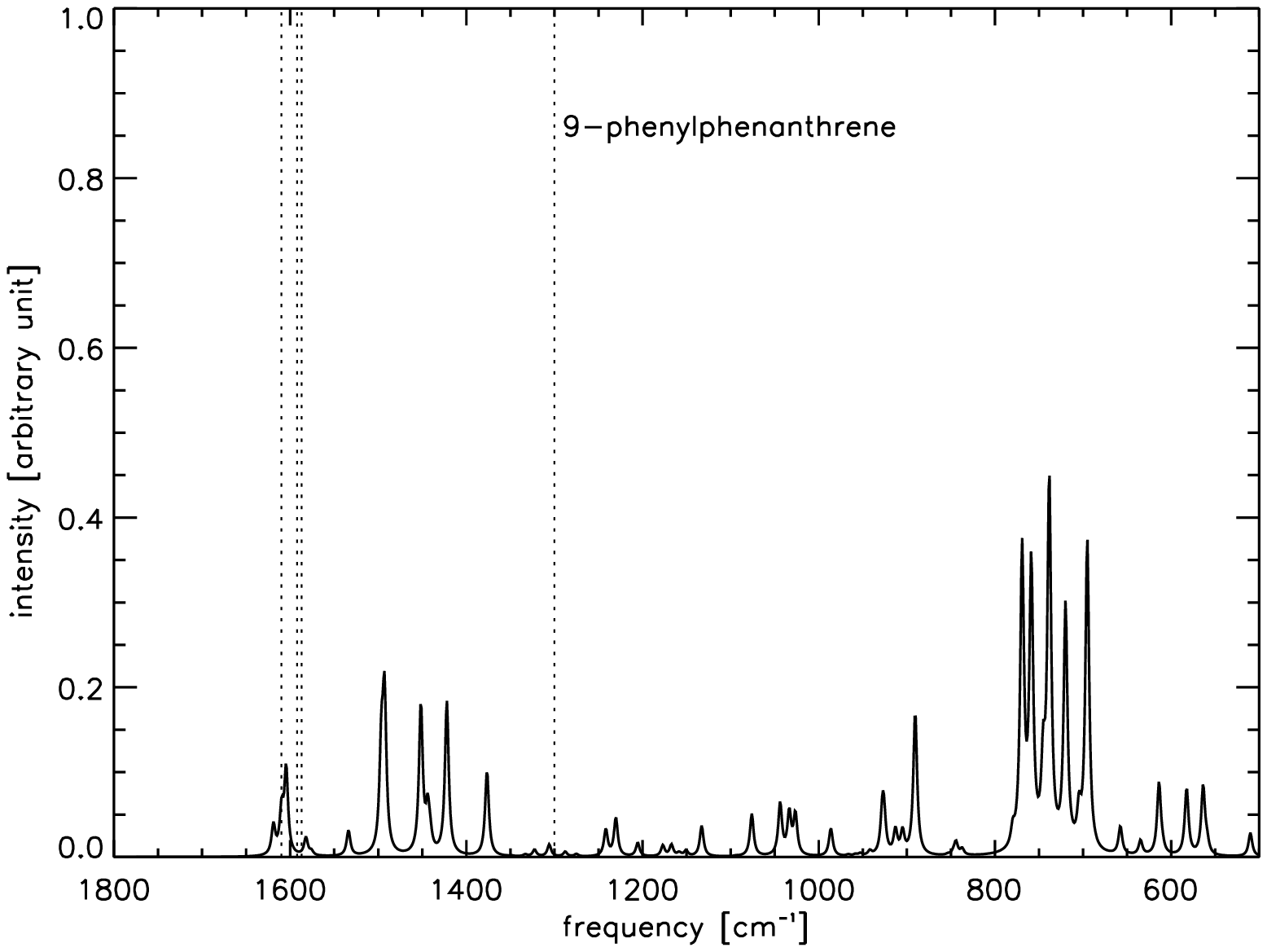}}
\centerline{\includegraphics[width=0.8\textwidth]{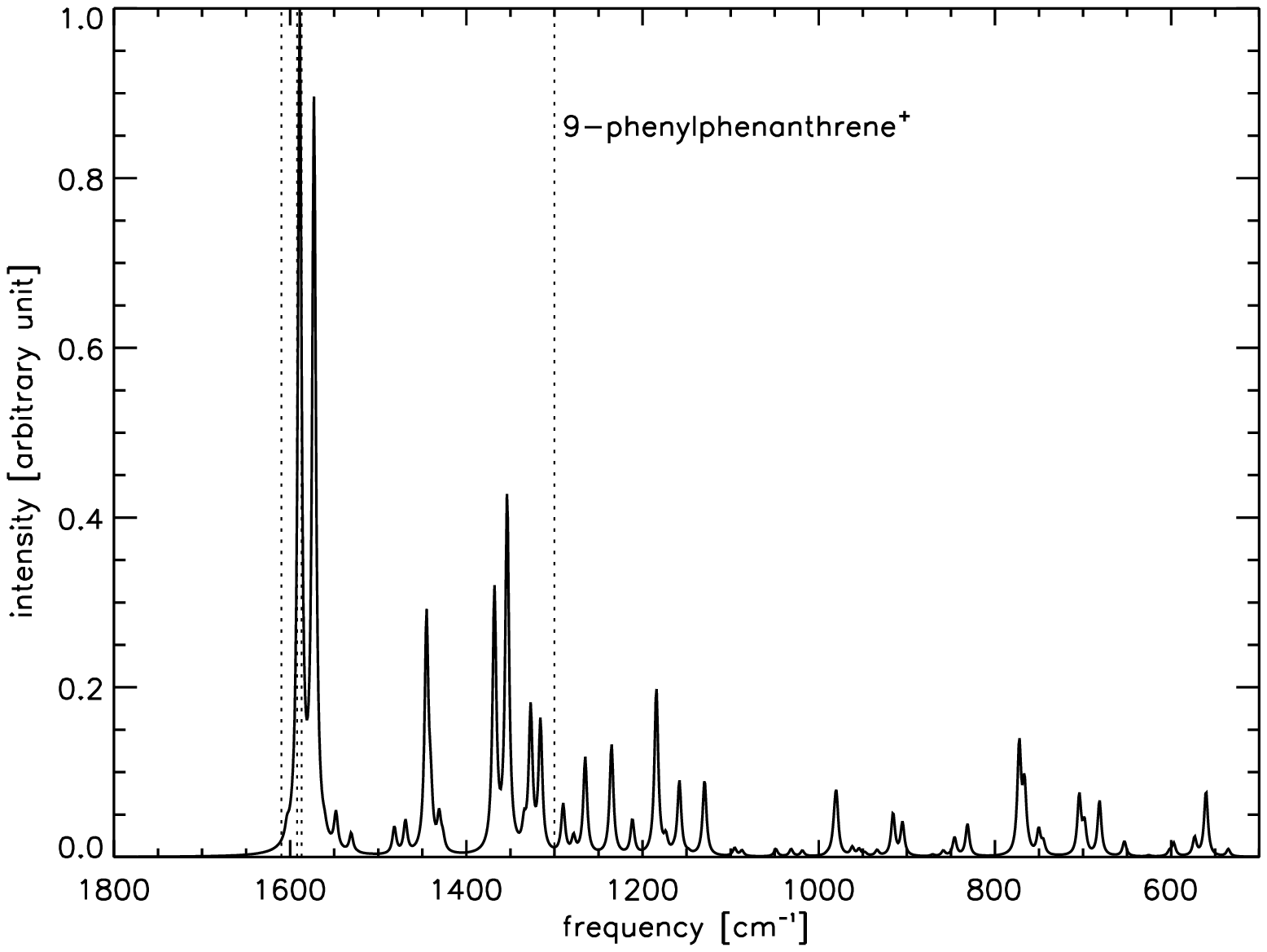}}
\caption{Infrared spectra of 9--phenylphenanthrene neutral and cation}
\label{Fig7-4}
\end{figure}

\clearpage

\begin{figure}
\vspace{-0.8in}
        \begin{tabular}{cc}
                      \includegraphics[width=0.4\textwidth]{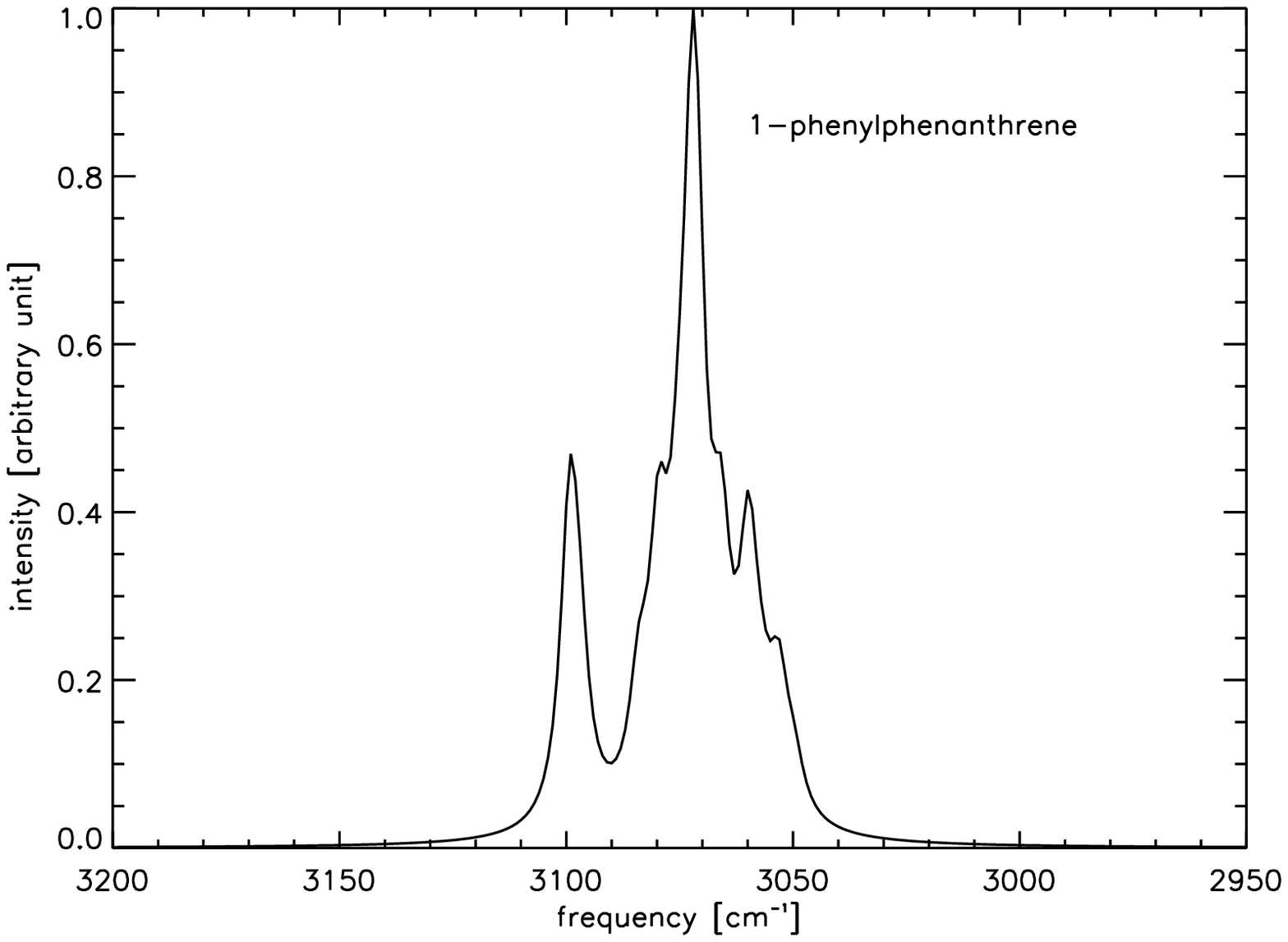}&
                      \includegraphics[width=0.4\textwidth]{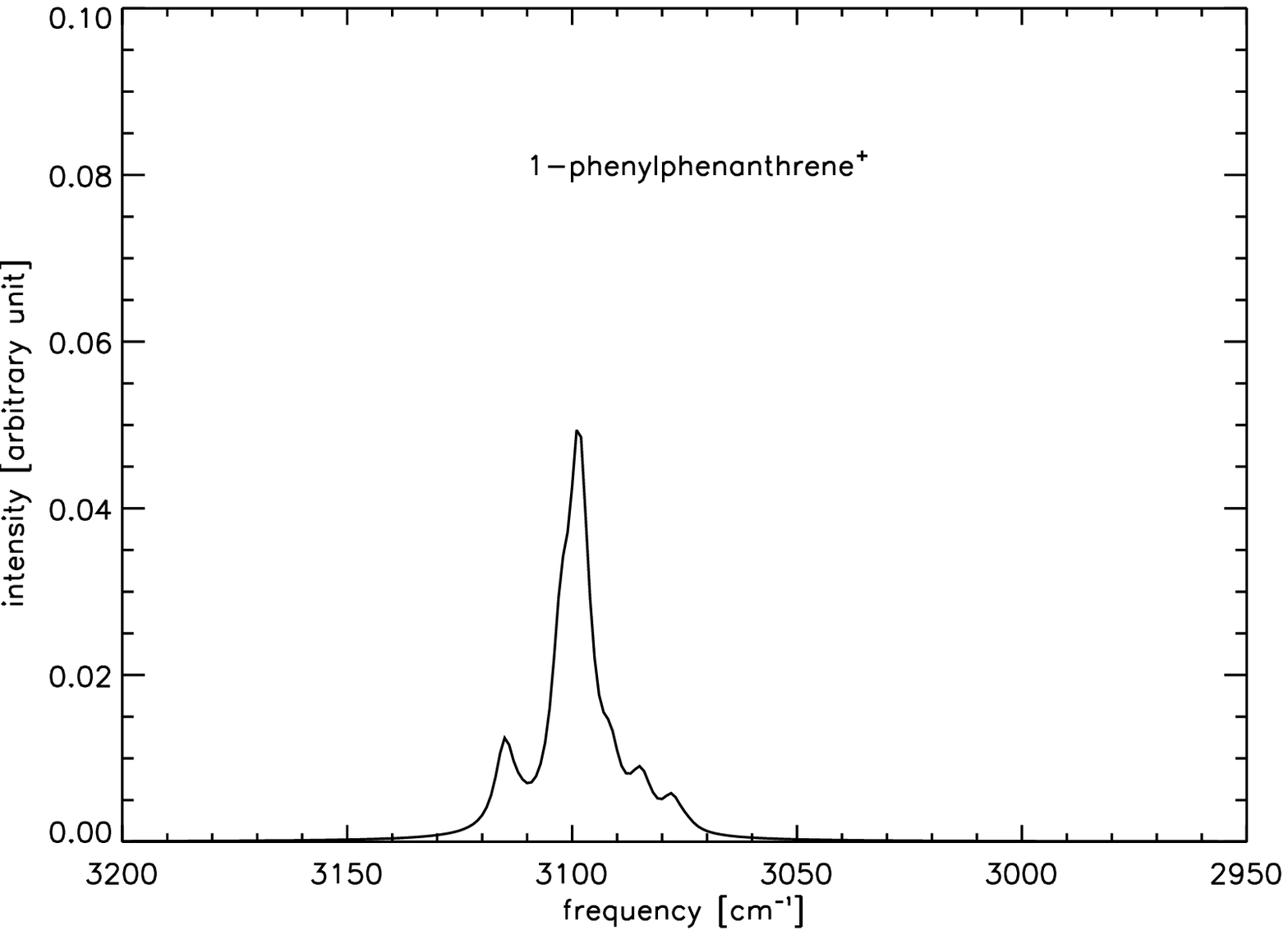}\\
                      \includegraphics[width=0.4\textwidth]{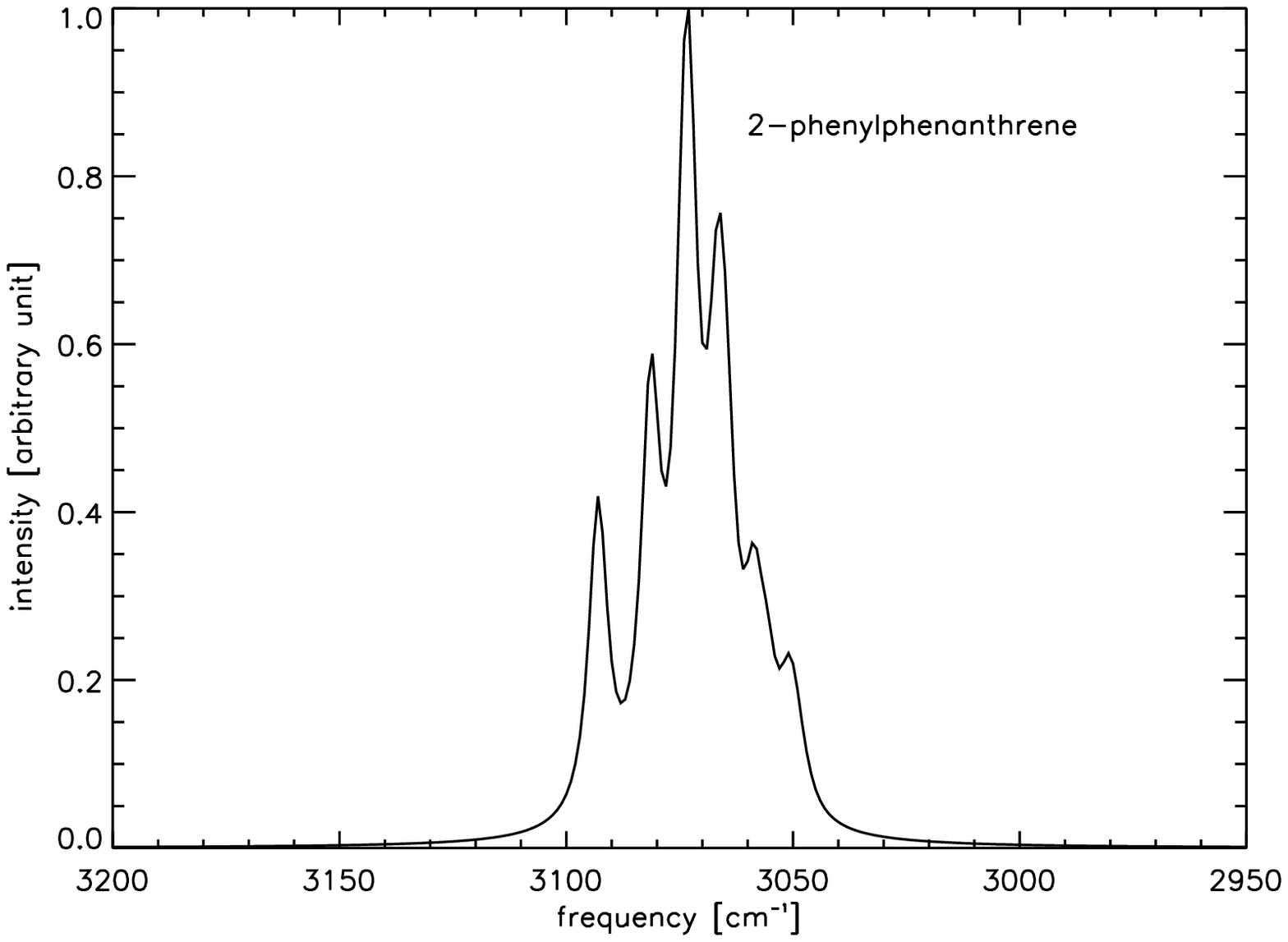}&
                      \includegraphics[width=0.4\textwidth]{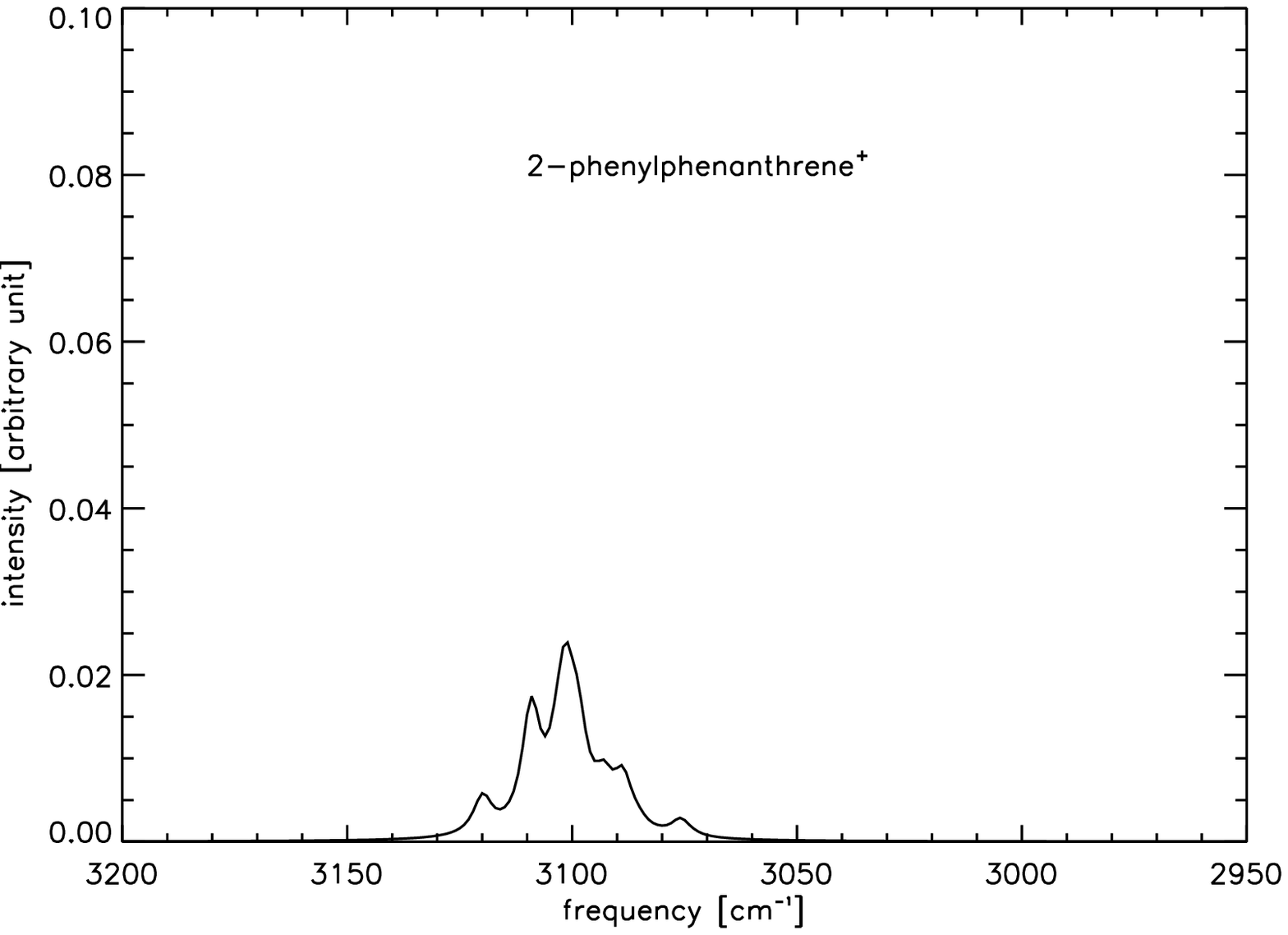}\\
                      \includegraphics[width=0.4\textwidth]{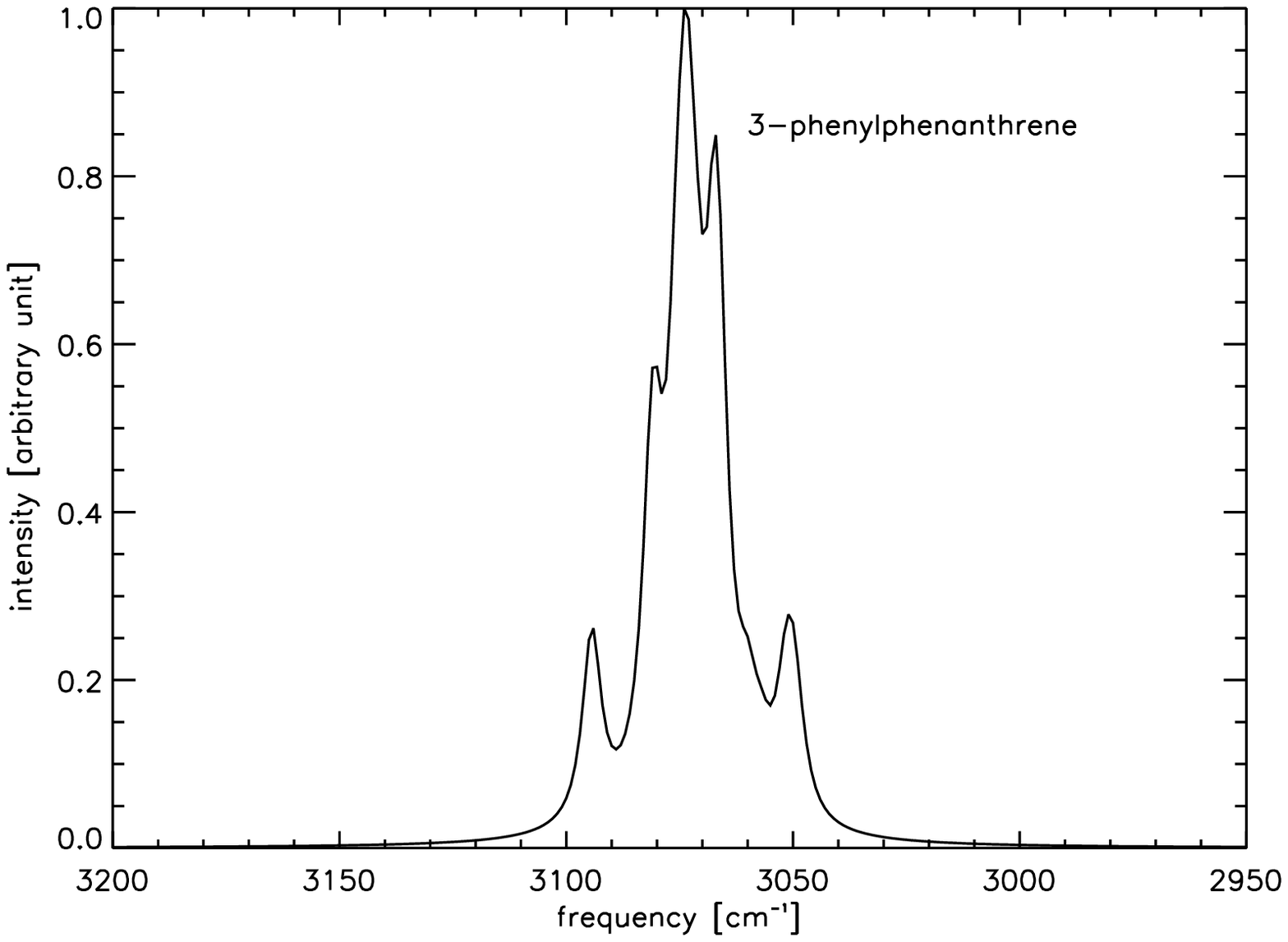}&
                      \includegraphics[width=0.4\textwidth]{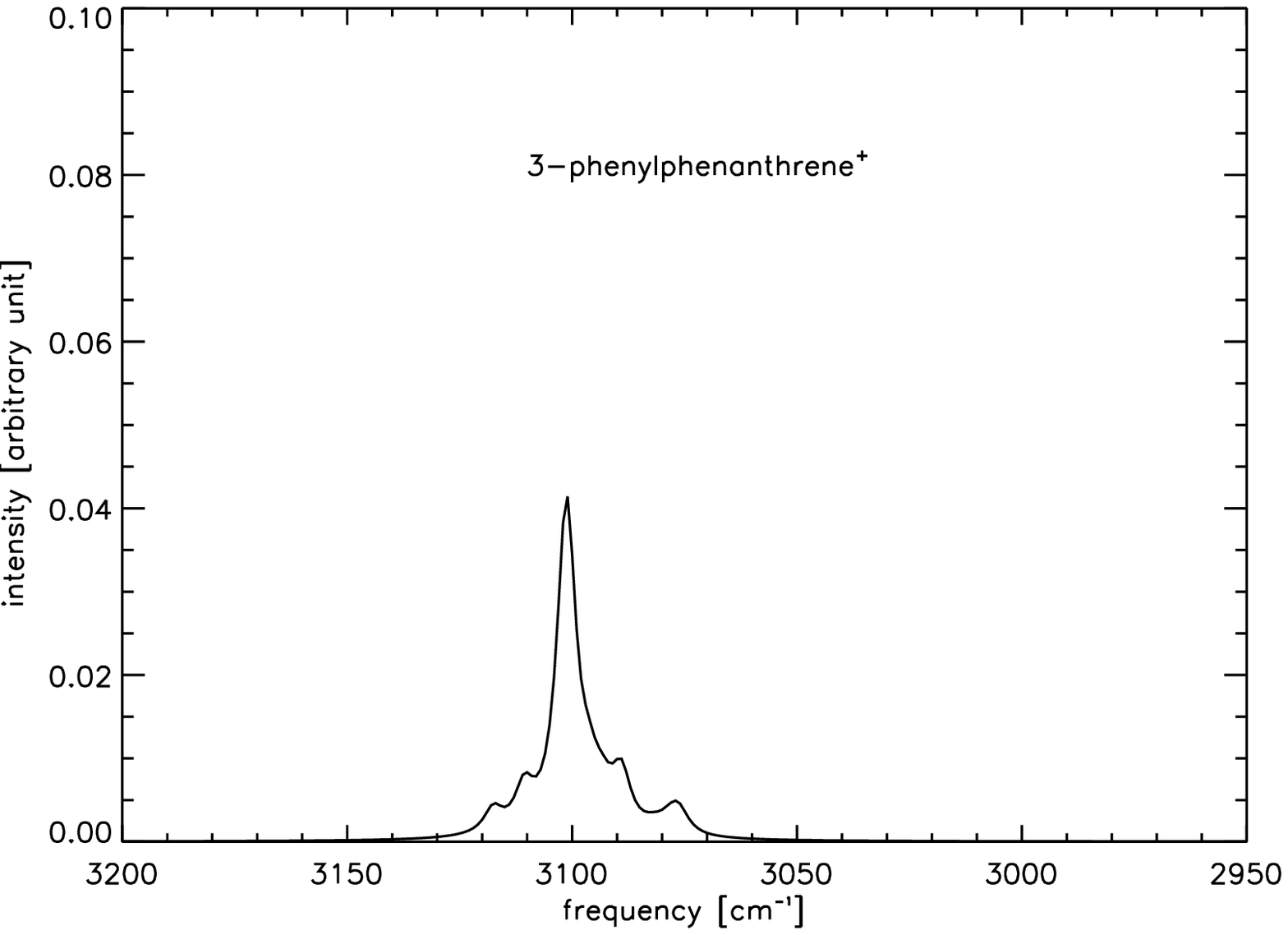}\\
                      \includegraphics[width=0.4\textwidth]{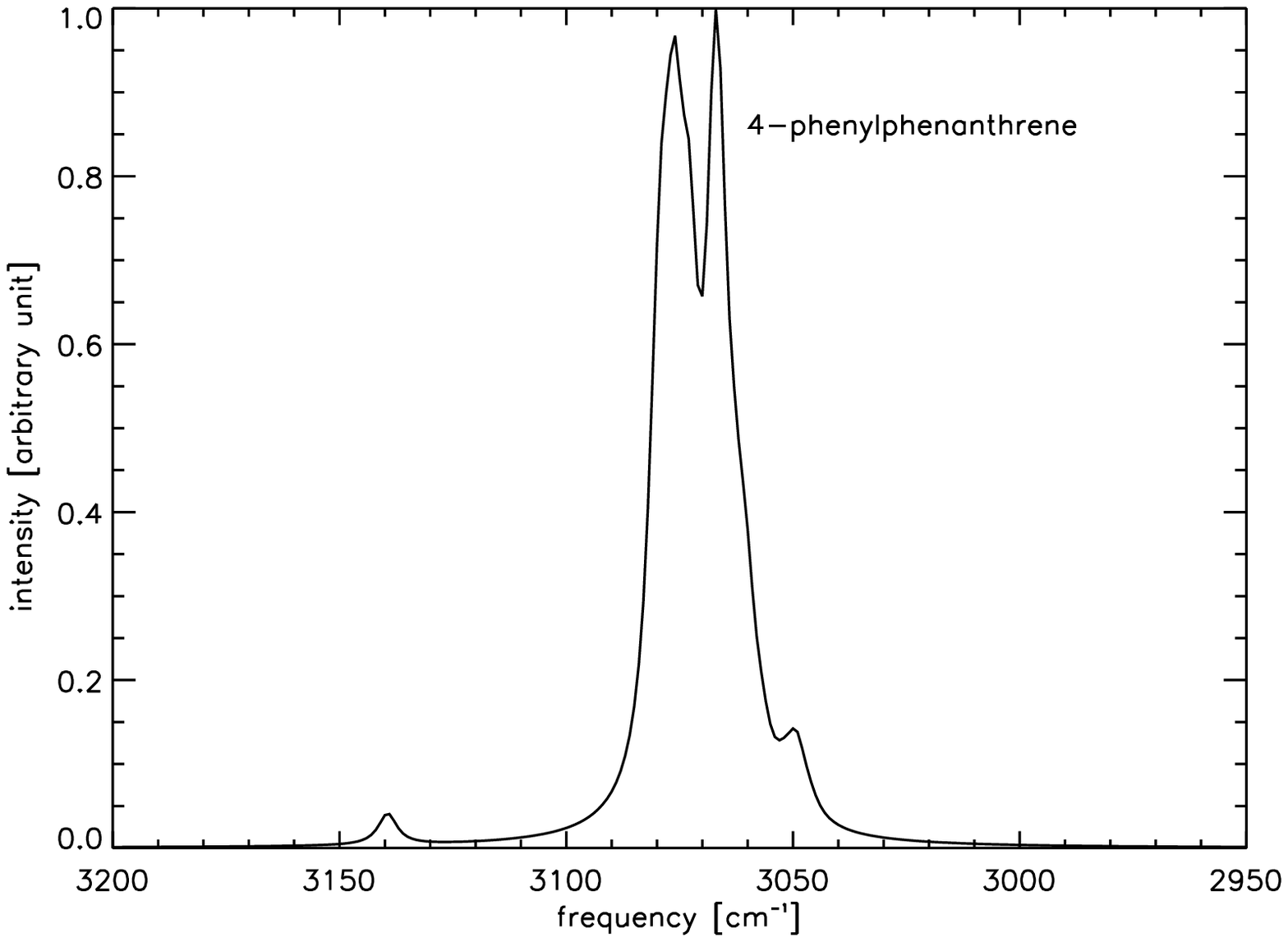}&
                      \includegraphics[width=0.4\textwidth]{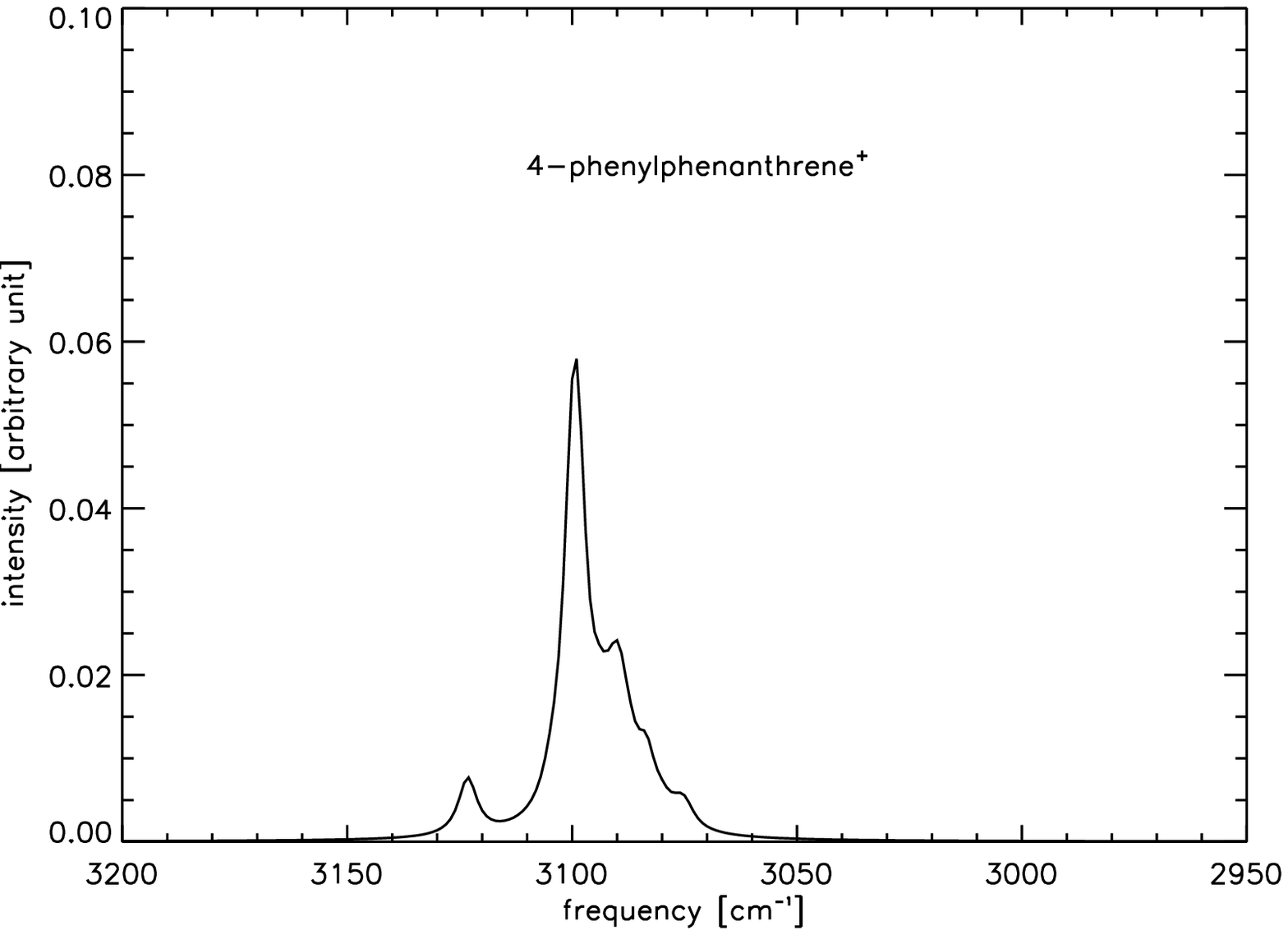}\\
                      \includegraphics[width=0.4\textwidth]{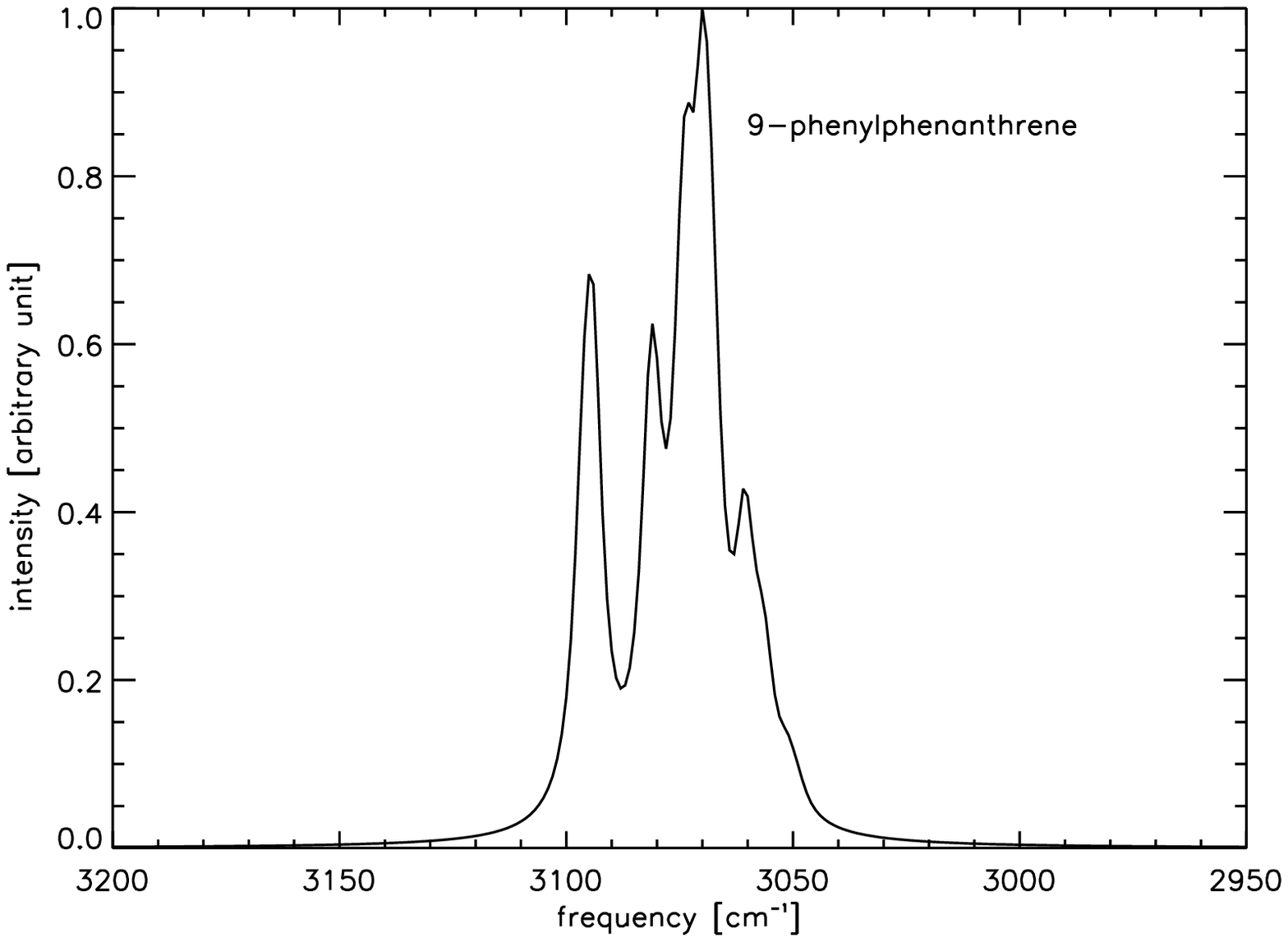}&
                      \includegraphics[width=0.4\textwidth]{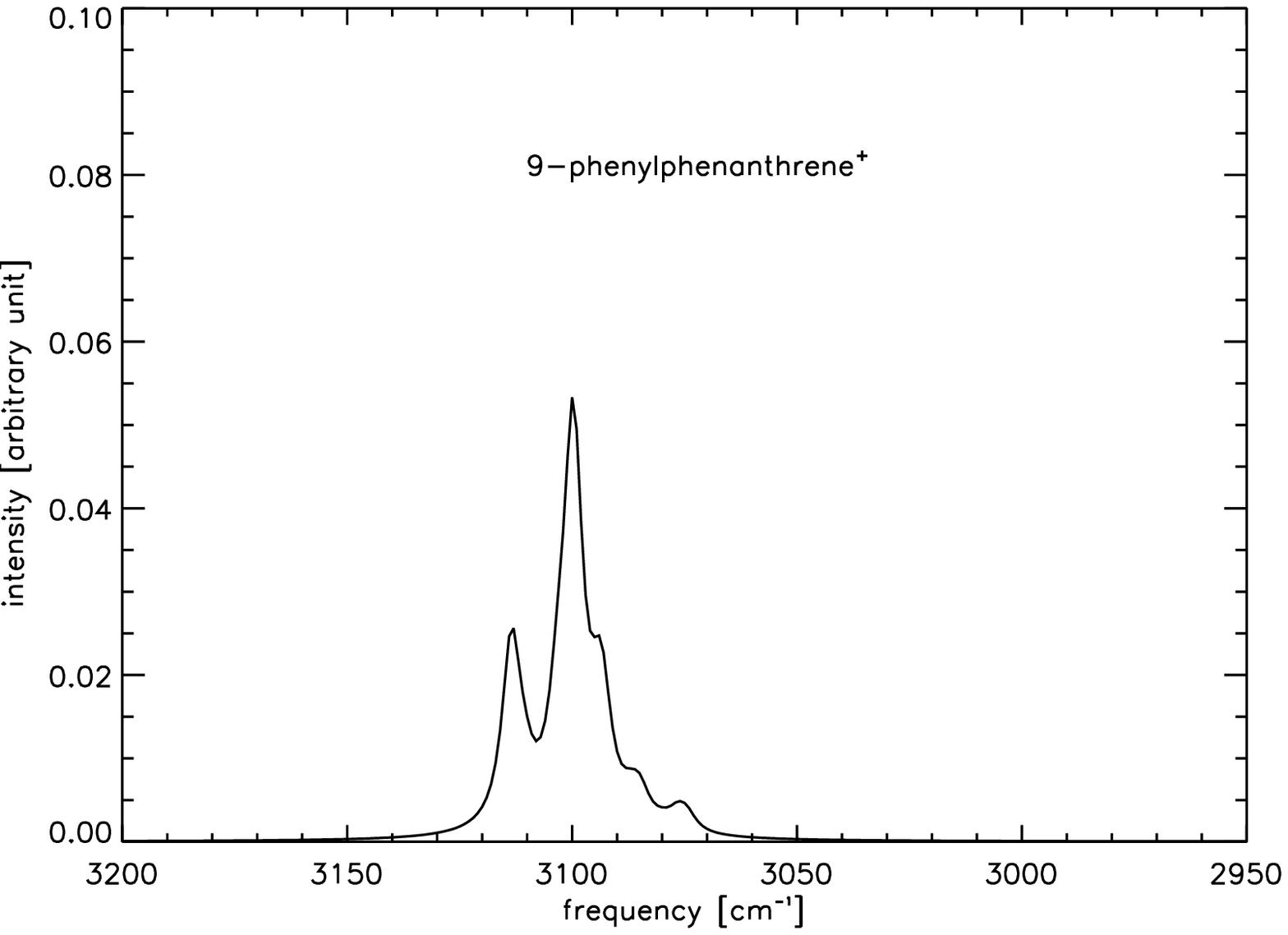}
        \end{tabular}
\caption{The C-H stretch spectral region of infrared spectra in neutrals and cations of phenylphenanthrenes.}
\label{Fig8-4}
\end{figure}



\clearpage

\begin{table}
\caption{Optimized geometry-phenyl group torsion angle and energies}
\label{tab1-4}
\begin{tabular}{cccccc}
\hline
\multicolumn{1}{c}{Phenylphenanthrenes}&\multicolumn{2}{c}{torsion angle}&\multicolumn{2}{c}{optimization energy}&\multicolumn{1}{c}{Energy change}\\
  & \multicolumn{2}{c}{(degree)}& \multicolumn{2}{c}{(Kcal/mol)}  &\multicolumn{1}{c}{$\Delta$~E}\\
      &\multicolumn{1}{c}{neutral}&\multicolumn{1}{c}{cation}&\multicolumn{1}{c}{neutral}&\multicolumn{1}{c}{cation}\\
\hline
1-Phenylphenanthrene&56.2&43.2&-770.0970402&-769.8434146&-0.2536\\
2-Phenylphenanthrene&36.9&23.6&-770.1019447&-769.8451057&-0.2568\\
3-Phenylphenanthrene&41.1&24.2&-770.1018530&-769.8451170&-0.2567\\
4-Phenylphenanthrene&61.3&43.3&-770.0886520&-769.8325988&-0.2561\\
9-Phenylphenanthrene&56.3&42.6&-770.0973570&-769.8451452&-0.2522\\
\hline  
\end{tabular}
\end{table}

\begin{table}
\caption{Infrared frequencies (cm$^{-1}$) and relative intensities for 1--phenylphenanthrene}
\label{tab2-4}
\tiny    
\begin{center}
\begin{tabular}{ccccccc}
\hline 
\multicolumn{7}{c}{1--phenylphenanthrene} \\
\multicolumn{2}{c}{Neutral}&\multicolumn{2}{c}{Cation}\\ \cline{1-4} \cline{5-7}
\multicolumn{1}{l}{Frequency}&\multicolumn{1}{c}{Relative Intensity}&       \multicolumn{1}{c}{Frequency}&\multicolumn{1}{c}{Relative Intensity} \\
\hline
99.91&0.01&118.68&0.01\\
216.07&0.03	&209.57&0.01\\
218.17&0.01	&214.67&0.01\\
300.14&0.02	&245.77&0.03\\
327.68&0.01	&402.45&0.04\\
401.16&0.02&440.5&0.01\\
407.2&0.01&465.87&0.02\\
412.65&0.02	&534.4&0.01\\
446.88&0.08	&544.43&0.03\\
490.21&0.02	&595.2&0.01\\
496.9&0.08&654.18&0.08\\
500.18&0.02	&682.67&0.04\\
543.66&0.03	&699.07&0.01\\
555.06&0.05	&715.36&0.02\\
570.09&0.01	&749.01&0.06\\
604.9&0.04&750.54&0.04\\
609.17&0.27&760.58&0.03\\
622.68&0.03&789.25&0.13\\
661.2&0.02&804.75&0.04\\
696.5&0.56&823.17&0.01\\
704.41&0.06	&833.58&0.04\\
711.03&0.01&841.1&0.02\\
742.2&0.70&856.77&0.01\\
746.22&0.75	&871.22&0.01\\
758.26&0.28	&967.75&0.01\\
794.41&0.02&975.15&0.02\\
802.62&0.49	&980.35&0.05\\
821.83&0.23	&983.08&0.01\\
826.73&0.08	&1018.96&0.01\\
838.1&0.01&1071.37&0.03\\
855.59&0.12&1078.28&0.01\\
864.37&0.10	&1090.29&0.01\\
882.79&0.03	&1137.58&0.03\\
906.42&0.03	&1145.46&0.30\\
914.77&0.03	&1166.61&0.07\\
938.14&0.02	&1182.51&0.14\\
940.95&0.02	&1189.15&0.08\\
959.99&0.01	&1225.1&0.04\\
985.14&0.03	&1245.09&0.28\\
990.95&0.04	&1258.47&0.03\\
1026.14&0.10&1268.11&0.03\\
1040.98&0.05&1291.59&0.01\\
1068.51&0.08&1309.36&0.14\\
1076.18&0.02&1317.35&0.03\\
1080.29&0.04&1334.1&0.01\\
1138.4&0.01	&1347.27&0.48\\
1167.53&0.06&1384.01&	0.01\\
1174.42&0.02&1419.12&0.18\\
1188.16&0.01&1430.17&0.17\\
1201.92&0.04&1437.51&0.03\\
1229.58&0.07&1446.48&0.21\\
1260.16&0.12&1480.93&0.13\\
1261.91&0.03&1503.48&0.06\\
1286.46&0.02&1510.92&0.12\\
1295.7&0.05	&1522.75&0.20\\
1319.37&0.01&1547.63&0.36\\
1347.9&0.01	&1568.01&0.02\\
1383.03&0.11&1586.82&1.00\\
1423.6&0.09	&1608.12&0.10\\
1439.08&0.09&3091.49&0.01\\
1441.83&0.11&3098.47&0.04\\
1461.13&0.25&3102.4&0.02\\
1493.23&0.11&3114.96&0.01\\
1503.04&0.15&	&&	\\
1530.44&0.11&	&&	\\
1574.53&0.06&	&&	\\
1578.47&0.06&	&&	\\
1598.72&0.13&	&&	\\
1608.28&0.13&	&&	\\
1617.38&0.04&	&&	\\
1627.14&0.01&	&&	\\
3048.94&0.01&	&&	\\
3049.81&0.08&	&&	\\
3053.16&0.24&	&&	\\
3056.08&0.04&	&&	\\
3057.45&0.08&	&&	\\
3059.81&0.49&	&&	\\
3065.83&0.43&	&&	\\
3071.05&0.53&	&&	\\
3072.25&1.00&	&&	\\
3075.27&0.31&	&&	\\
3079.62&0.47&	&&	\\
3083.95&0.22&	&&	\\
3097.08&0.23&	&&	\\
3099.2&0.63	&	&&       \\
\hline
\multicolumn{5}{p{0.6\textwidth}}{The relative intensities are obtained by normalizing with the strongest band of each set. Maximum absolute intensity: neutral, 1.12 Debye$^{2}$/AMU \AA$^{2}$; cation, 14.10 Debye$^{2}$/AMU \AA$^{2}$.}
\end{tabular}
\end{center}
\end{table}

\begin{table}
\caption{Infrared frequencies (cm$^{-1}$) and relative intensities for 2--phenylphenanthrene}
\label{tab3-4}
\tiny
\begin{center}
\begin{tabular}{ccccccc}
\hline 
\multicolumn{7}{c}{2--phenylphenanthrene} \\
\multicolumn{2}{c}{Neutral}&\multicolumn{2}{c}{Cation}\\ \cline{1-4} \cline{5-7}
\multicolumn{1}{l}{Frequency}&\multicolumn{1}{c}{Relative Intensity}&\multicolumn{1}{c}{Frequency}&\multicolumn{1}{c}{Relative Intensity} \\
\hline
216.1&0.01&410.93&0.01\\
226.99&0.04&443.15&0.01\\
391.22&0.01	&455.03&0.03\\
430.49&0.11	&464.61&0.01\\
446.38&0.13	&496.99&0.01\\
459.55&0.01	&538.3&0.01\\
488.19&0.03	&582.51&0.01\\
513.61&0.05&659.69&0.02\\
531.58&0.01	&660.61&0.03\\
535.71&0.05	&698.15&0.02\\
547.51&0.03	&735.1&0.01\\
629.01&0.01	&743.74&0.02\\
672.37&0.33	&758.98&0.06\\
692.97&0.49	&796.76&0.03\\
705.52&0.05	&865.78&0.01\\
713.91&0.06	&869.08&0.02\\
737.25&0.76	&898.24&0.02\\
744.66&0.28	&978.32&0.07\\
753.9&0.54&1006.24&0.03\\
772.67&0.03	&1029.72&0.01\\
803.23&0.76	&1088.06&0.06\\
818.34&0.22	&1119.08&0.15\\
835.09&0.01	&1155.04&0.02\\
844.95&0.09	&1163.38&0.01\\
853.76&0.04	&1165.2&0.01\\
881.76&0.13	&1192.86&0.12\\
883.15&0.29	&1197.72&0.01\\
904.23&0.02	&1205.79&0.01\\
912.49&0.06	&1238.88&0.04\\
930.4&0.02&1253.2&0.01\\
940.72&0.01	&1272.58&0.04\\
983.28&0.08	&1305.26&0.02\\
987.3&0.01&1312.25&0.09\\
1020.05&0.09&1332.15&0.13\\
1035.87&0.05&1343.27&0.01\\
1045.8&0.01	&1351.62&0.12\\
1080.04&0.07&1398.92&0.03\\
1094.55&0.01&1410.71&0.04\\
1145.08&0.03&1420.15&0.06\\
1162.84&0.01&1441.27&0.03\\
1180.91&0.05&1449.27&0.07\\
1186.33&0.07&1461.94&0.04\\
1204.91&0.02&1494.1&0.07\\
1237.93&0.15&1515.57&0.03\\
1266.29&0.07&1583.36&1.00\\
1286.68&0.05&1595.2&0.26\\
1305.32&0.01&1599.03&0.02\\
1325.84&0.01&1622.01&0.03\\
1343.21&0.01&3088.69&0.01\\
1345.87&0.01&3098.71&0.01\\
1383.35&0.05&3101.52&0.02\\
1421.77&0.06&3109.09&0.01\\
1427.34&0.04&	&&	\\
1447.47&0.08&	&&	\\
1464.13&0.51&	&&	\\
1492.25&0.48&	&&	\\
1501.51&0.08&	&&	\\
1531.19&0.01&	&&	\\
1567.46&0.02&	&&	\\
1585.34&0.03&	&&	\\
1606.62&0.11&	&&	\\
1611.06&0.23&	&&	\\
1619.15&0.06&	&&	\\
1624.73&0.19&	&&	\\
3046.91&0.01&	&&	\\
3049.17&0.11&	&&	\\
3050.97&0.20&	&&	\\
3055.62&0.18&	&&	\\
3058.19&0.15&	&&	\\
3058.9&0.21	&	&&         \\
3064.62&0.40&	&&	\\
3066.47&0.76&	&&	\\
3068.59&0.10&	&&	\\
3072.92&1.00&	&&	\\
3074.13&0.63&	&&	\\
3080.58&0.08&	&&	\\
3081.44&0.76&	&&	\\
3092.97&0.67&	&&	\\
\hline
\multicolumn{5}{p{0.6\textwidth}}{The relative intensities are obtained by normalizing with the strongest band of each set. Maximum absolute intensity: neutral, 1.04 Debye$^{2}$/AMU \AA$^{2}$; cation, 27.03 Debye$^{2}$/AMU \AA$^{2}$.}
\end{tabular}
\end{center}
\end{table}

\begin{table}
\vspace{-1.25in}
\caption{Infrared frequencies (cm$^{-1}$) and relative intensities for 3--phenylphenanthrene}
\label{tab4-4}
\tiny
\begin{center}
\begin{tabular}{ccccccc}
\hline 
\multicolumn{7}{c}{3--phenylphenanthrene} \\
\multicolumn{2}{c}{Neutral}&\multicolumn{2}{c}{Cation}\\ \cline{1-4} \cline{5-7}
\multicolumn{1}{l}{Frequency}&\multicolumn{1}{c}{Relative Intensity}&\multicolumn{1}{c}{Frequency}&\multicolumn{1}{c}{Relative Intensity} \\
\hline
106.26&0.01	&262.31&0.01\\
112.83&0.01	&391.94&0.01\\
217.78&0.01	&411.89&0.01\\
304.56&0.02	&504.93&0.01\\
365.47&0.02	&562.74&0.02\\
409.19&0.02	&596.44&0.01\\
416.52&0.02	&607.84&0.01\\
427.51&0.08	&618.06&0.01\\
453.77&0.03	&667.52&0.02\\
505.79&0.03	&677.33&0.03\\
512.93&0.09	&706.69&0.02\\
525.5&0.04&738.03&0.03\\
552.72&0.03&746.76&0.04\\
566.03&0.11&765.74&0.03\\
611.25&0.02	&820.57&0.01\\
620.81&0.10	&844.55&0.02\\
626.59&0.07	&850.75&0.03\\
680.11&0.05	&859.84&0.02\\
692.95&0.42	&870.97&0.02\\
705.78&0.06	&880.91&0.01\\
709.41&0.09	&978.38&0.06\\
736.12&0.64	&984.34&0.01\\
738.51&0.29	&1026.36&0.02\\
755.59&0.42	&1093.65&0.01\\
777.27&0.01	&1095.31&0.02\\
794.49&0.21	&1138.71&0.10\\
834.87&0.02	&1159.87&0.01\\
837.38&0.87	&1164.14&0.01\\
848.64&0.05	&1192.55&0.20\\
856.34&0.24	&1208.1&0.04\\
865.03&0.11&1221.71&0.04\\
869.79&0.06&1247.25&0.11\\
903.26&0.02&1260.16&0.08\\
917.34&0.06&1277.91&0.01\\
934.3&0.01&1316.14&0.10\\
940.48&0.01&1332.53&0.12\\
985.84&0.01&1344.17&0.04\\
991.61&0.06&1356.02&0.07\\
1023.83&0.03&1408.59&0.10\\
1034.71&0.10&1418.79&0.07\\
1044.11&0.02&1443.98&0.03\\
1078.89&0.07&1453.51&0.10\\
1094.77&0.04&1476.23&0.01\\
1146.01&0.03&1491.97&0.05\\
1160.46&0.02&1524.57&0.21\\
1180.15&0.03&1549.71&0.01\\
1197.27&0.02&1556.02&1.00\\
1214.56&0.01&1583.69&0.21\\
1227.45&0.19&1596.86&0.05\\
1250.07&0.07&1605.1&0.17\\
1269.44&0.01&3089.01&0.01\\
1280.59&0.05&3101.15&0.03\\
1302.51&0.01&3101.71&0.01\\
1325.08&0.02&	       &      \\
1340.14&0.01&	       &     \\
1349.38&0.04&	       &     \\
1396.19&0.10&	       &	    \\
1421.77&0.03&	      &	   \\
1426.28&0.02&	      &      \\
1448.49&0.12&	     &	   \\
1454.08&0.28&	    &	  \\
1496.6&0.45	&	    &         \\
1511.56&0.15&	   &	  \\
1519.11&0.05&	   &	\\
1565.62&0.02&	   &	\\
1585.49&0.03&	&	\\
1609.25&0.07&	&	\\
1612.13&0.35&	&	\\
1618.28&0.08&	&	\\
1623.18&0.05&	&	\\
3047.38&0.01&	&	\\
3049.62&0.20&	&	\\
3050.92&0.13&	&	\\
3051.84&0.14&	&	\\
3057.02&0.08&	&	\\
3060.03&0.16&	&	\\
3066.36&0.62&	&	\\
3067.47&0.60&	&	\\
3071.31&0.40&	&	\\
3073.56&1.00&	&	\\
3075.74&0.53&	&	\\
3080.14&0.15&	&	\\
3081.07&0.59&	&	\\
3094.36&0.40&	&	\\
\hline
\multicolumn{5}{p{0.6\textwidth}}{The relative intensities are obtained by normalizing with the strongest band of each set. Maximum absolute intensity: neutral, 1.11 Debye$^{2}$/AMU \AA$^{2}$; cation, 21.68 Debye$^{2}$/AMU \AA$^{2}$.}
\end{tabular}
\end{center}
\end{table}

\begin{table}
\caption{Infrared frequencies (cm$^{-1}$) and relative intensities for 4--phenylphenanthrene}
\label{tab5-4}
\tiny
\begin{center}
\begin{tabular}{ccccccc}
\hline 
\multicolumn{7}{c}{4--phenylphenanthrene} \\
\multicolumn{2}{c}{Neutral}&\multicolumn{2}{c}{Cation}\\ \cline{1-4} \cline{5-7}
\multicolumn{1}{l}{Frequency}&\multicolumn{1}{c}{Relative Intensity}&\multicolumn{1}{c}{Frequency}&\multicolumn{1}{c}{Relative Intensity} \\
\hline
103.55&0.02	&49.88&0.01\\
198.3&0.01&126.36&0.04\\
220.05&0.02	&182.34&0.04\\
230.13&0.02	&260.38&0.02\\
306.07&0.01	&290.88&0.01\\
329.2&0.01&330&0.04\\
402.53&0.01	&376.86&0.20\\
405.43&0.01	&448.43&0.03\\
409.08&0.01	&476.08&0.07\\
455.42&0.01	&486.5&0.01\\
478.4&0.01&506.18&0.06\\
494.84&0.02	&532.69&0.03\\
508&0.03&534.84&0.01\\
535.67&0.13&654.66&0.07\\
556.62&0.12	&677.15&0.06\\
605.27&0.05	&695.53&0.02\\
609.19&0.11	&701.83&0.01\\
625.71&0.01	&745.34&0.06\\
662.79&0.05	&747.45&0.03\\
695.77&0.55	&756.31&0.06\\
705.58&0.04	&763.37&0.03\\
713.46&0.21	&792.66&0.01\\
736.29&0.86	&818.99&0.13\\
757.62&0.32	&828.48&0.06\\
761.73&0.07	&831.86&0.05\\
767.48&0.09	&864.31&0.04\\
788.29&0.16	&900.27&0.10\\
821.82&0.67	&931.33&0.01\\
826.49&0.33	&947.62&0.01\\
857.59&0.11	&959.83&0.01\\
887.81&0.02	&969.69&0.04\\
899.07&0.03	&976.06&0.01\\
912.25&0.06	&980.48&0.09\\
924.91&0.04	&1062.28&0.02\\
936.43&0.02&1086.17&0.01\\
958.7&0.02&1094.02&0.10\\
980.86&0.04&1121.79&0.37\\
985.35&0.02&1138.84&0.17\\
1020.04&0.08&1180.84&0.02\\
1040.33&0.04&1183.66&0.31\\
1064.92&0.04&1205.86&0.08\\
1071.19&0.06&1218.94&0.01\\
1087.53&0.04&1241.99&0.04\\
1135.03&0.05&1256.19&0.49\\
1160.46&0.10&1293.99&0.02\\
1171.76&0.01&1305.68&0.15\\
1175.03&0.03&1326.95&0.30\\
1199.59&0.02&1337.81&0.11\\
1212.96&0.03&1343.35&0.03\\
1248.02&0.05&1380.65&0.03\\
1260.47&0.06&1399.54&1.00\\
1282.94&0.03&1420.5&0.01\\
1291.56&0.13&1444.5&0.15\\
1312.41&0.02&1450.48&0.06\\
1318.69&0.02&1483.95&0.02\\
1383.75&0.09&1498.31&0.07\\
1435.11&0.21&1519.86&0.07\\
1437.91&0.12&1540.92&0.38\\
1448.96&0.20&1548.08&0.01\\
1491.26&0.20&1565.16&0.59\\
1498.04&0.03&1591.73&0.23\\
1526.71&0.03&1603.69&0.05\\
1569.24&0.04&3083.52&0.01\\
1579.61&0.05&3089.56&0.01\\
1595.93&0.04&3091.38&0.01\\
1605.17&0.14&3093.95&0.01\\
1615.2&0.05	&3098.75&0.03\\
1628.2&0.02	&3099.94&0.03\\
3046.37&0.02&3123.2&0.01\\
3048.92&0.09&           &	    \\
3049.86&0.05&	    &         \\
3052&	0.04&	                &       \\
3056.97&0.07&	     &      \\
3060.34&0.28&	    &        \\
3062.93&0.26&	    &        \\
3066.09&0.47&	   &          \\
3067.2&1.00	&	   &          \\
3072.9&0.74	&	&              \\
3076.09&0.90&	&	      \\
3078.82&0.67&	&	        \\
3080.38&0.30&	&	       \\
3139.4&0.07	&	&               \\
\hline
\multicolumn{5}{p{0.6\textwidth}}{The relative intensities are obtained by normalizing with the strongest band of each set. Maximum absolute intensity: neutral, 1.08 Debye$^{2}$/AMU \AA$^{2}$; cation,12.82 Debye$^{2}$/AMU \AA$^{2}$.}
\end{tabular}
\end{center}
\end{table}

\begin{table}
\vspace{-0.5in}
\caption{Infrared frequencies (cm$^{-1}$) and relative intensities for 9--phenylphenanthrene}
\label{tab6-4}
\tiny
\begin{center}
\begin{tabular}{ccccccc}
\hline 
\multicolumn{7}{c}{9--phenylphenanthrene}\\ 
\multicolumn{2}{c}{Neutral}&\multicolumn{2}{c}{Cation}\\ \cline{1-4} \cline{5-7}
\multicolumn{1}{l}{Frequency}&\multicolumn{1}{c}{Relative Intensity}&\multicolumn{1}{c}{Frequency}&\multicolumn{1}{c}{Relative Intensity} \\
\hline
103.21&0.01	&233.15&0.02\\
202.33&0.01	&259.32&0.01\\
237.29&0.01	&320.87&0.01\\
256.55&0.01	&389.08&0.03\\
277.14&0.02	&399.91&0.02\\
401.7&0.03&422.33&0.01\\
406.15&0.01&475.84&0.01\\
427.07&0.07&535.06&0.01\\
455.19&0.01&560.45&0.08\\
479.14&0.02&573.36&0.02\\
510.01&0.05&597.05&0.02\\
559.34&0.02&601.79&0.01\\
563.73&0.14&653.02&0.02\\
582.42&0.13&681.26&0.07\\
609.89&0.01&698.4&0.04\\
613.76&0.15&704.31&0.07\\
634.71&0.03&745.09&0.01\\
657.59&0.06&750.3&0.03\\
695.25&0.63&766.45&0.08\\
704.53&0.07&772.38&0.13\\
719.88&0.49&831.19&0.04\\
738.35&0.74&845.86&0.02\\
745.2&0.15&858.63&0.01\\
758.74&0.56&904.93&0.04\\
769.15&0.60&915.59&0.05\\
779.71&0.03&933.76&0.01\\
837.24&0.01&954.26&0.01\\
844.3&0.03&961.97&0.01\\
890.49&0.29&979.81&0.05\\
904.85&0.05&981.35&0.03\\
912.89&0.05&983.19&0.01\\
925.88&0.07&1018.55&0.01\\
927.66&0.08&1031.35&0.01\\
941.93&0.01&1048.69&0.01\\
986.18&0.05&1087.2&0.01\\
986.64&0.01&1095.48&0.01\\
1026.47&0.08&1129.61&0.09\\
1033.42&0.08&1158.19&0.09\\
1043.79&0.10&1173.54&0.02\\
1076.01&0.09&1184.3&0.20\\
1132.89&0.06&1211.52&0.04\\
1150.59&0.01&1235.2&0.13\\
1158.67&0.01&1265.09&0.12\\
1167.17&0.02&1278.29&0.02\\
1176.89&0.02&1290.09&0.06\\
1205.46&0.03&1315.85&0.15\\
1230.3&0.08	&1326.97&0.17\\
1241.7&0.05	&1334.41&0.03\\
1287.73&0.01&1353.73&0.43\\
1305.89&0.02&1368.15&0.31\\
1322.56&0.01&1426.71&0.02\\
1376.6&0.17	&1430.84&0.04\\
1422.16&0.31&1440.55&0.06\\
1441.13&0.03&1445.19&0.28\\
1443.88&0.08&1469.16&0.04\\
1451.6&0.30&1481.77&0.03\\
1492.93&0.30&1530.7&0.02\\
1496.5&0.19&1547.97&0.04\\
1533.79&0.05&1560.87&0.01\\
1575.79&0.01&1572.89&0.89\\
1582.06&0.04&1589.14&1.00\\
1604.57&0.17&1603.57&0.02\\
1609.76&0.08&3093.31&0.01\\
1618.99&0.06&3099.73&0.05\\
3049.12&0.04&3103.02&0.01\\
3051.31&0.08&3113.49&0.02\\
3056.06&0.16&	      &&         \\
3056.81&0.07&	      &&        \\
3058.57&0.07&	      &&         \\
3060.66&0.45&	      &&          \\
3067.77&0.53&	      &&          \\
3069.9&1.00	&	      &&          \\
3073.58&0.72&	      &&	 \\   
3074.89&0.33&	      &&            \\
3080.68&0.50&	      &&	 \\
3081.77&0.36&	      &&	 \\
3094.14&0.83&	      &&	  \\  
3096.26&0.45&	      &&	 \\
\hline
\multicolumn{5}{p{0.6\textwidth}}{The relative intensities are obtained by normalizing with the strongest band of each set. Maximum absolute intensity: neutral, 0.97 Debye$^{2}$/AMU \AA$^{2}$; cation, 8.59 Debye$^{2}$/AMU \AA$^{2}$.}
\end{tabular}
\end{center}
\end{table}

\begin{table}
\caption{Prominent frequencies corresponding to modes involving motion of hydrogens in neutrals. The values in parenthesis are relative intensities.}
\label{tab7-4}
\footnotesize
\begin{center}
\begin{tabular}{lcccc}
\hline
\multicolumn{1}{l}{phenylpenanthrenes}&\multicolumn{1}{c}{C-H wag}&\multicolumn{1}{c}{C-H wag}&\multicolumn{1}{c}{C-H wag}&\multicolumn{1}{c}{C-H stretch}\\
\multicolumn{1}{c}{neutral}&\multicolumn{1}{c}{Phenyl}&\multicolumn{1}{c}{phenanthrene unit}&\multicolumn{1}{c}{phenanthrene unit}& \\
\hline
1--phenylphenanthrene&697 (0.56)&742 (0.70)&803 (0.49)&3072 (1.0) \\
 &     &  746 (0.75) &  & 3099 (0.63)\\
2--phenylphenanthrene&693 (0.49)&737 (0.76)& 803 (0.76) & 3073( 1.0) \\
 &     &  754 (0.54)&  & 3081 (0.8), 3092 (0.67) \\
3--phenylphenanthrene&693 (0.42)&736 (0.64) & 837 (0.87) & 3074 (1.0) \\
 &     &  756 (0.42) & & 3081 (0.6), 3094 (0.4) \\
4--phenylphenanthrene&696 (0.55)&736 (0.86) & 822 (0.67) & 3067 (1.0), 3076 (0.90) \\
 &     &  758 (0.32) & & 3079 (0.67)\\
9--phenylphenanthrene&695 (0.63)&738 (0.74)& 769 (0.6) &3070 (1.0) \\
 &     &  759 (0.56) & & 3094 (0.83)\\
\hline  

\end{tabular}
\end{center}
\end{table}

\begin{table}
\caption{The active planar ring deformation modes in neutral and cations. The values in parenthesis are relative intensities.}
\label{tab8-4}
\footnotesize
\begin{center}
\begin{tabular}{lccc}
\hline
\multicolumn{1}{l}{phenylpenanthrenes}&\multicolumn{2}{c}{Neutral}&\multicolumn{1}{c}{Cation}\\
\multicolumn{1}{l}~&\multicolumn{1}{c}{C-C st. + C-H in-plane}&\multicolumn{1}{c}{C-C stretch}&\multicolumn{1}{c}{C-C stretch}\\
\hline
1--phenylphenanthrene&1461 (0.25), 1503 (0.15) &1599 (0.13), 1608 (0.13) &1587 (1.0)\\
2--phenylphenanthrene&1464 (0.51), 1492 (0.48) &1606 (0.11), 1611 (0.23) &1583 (1.0)\\
3--phenylphenanthrene&1454 (0.28), 1497 (0.45) &1612 (0.35)     & 1556 (1.0) \\
4--phenylphenanthrene&1449 (0.20), 1491 (0.20) &1605 (0.14)     & 1565 (0.59), 1592 (0.23)\\
9--phenylphenanthrene&1451 (0,30), 1493 (0.30) &1605 (0.17)     & 1589 (1.0),  1604 (0.17)\\
\hline  

\end{tabular}
\end{center}
\end{table}

\end{document}